\documentclass[onecolumn,aip]{revtex4-1}
\usepackage{amssymb,amsmath,bm}
\usepackage{graphicx}%
\usepackage{dcolumn}
\usepackage{color}
\usepackage{makecell}
\usepackage{longtable}
\usepackage[dvips]{epsfig}
\usepackage[table, dvipsnames]{xcolor}
\usepackage{colortbl}
\usepackage[perpage,symbol*]
{footmisc}
\usepackage{framed}
\usepackage{multirow}
\usepackage{verbatim}
\usepackage{tablefootnote,threeparttable}
\usepackage{braket}
\usepackage{comment}
\usepackage{simpler-wick} 

\usepackage{lineno}

\usepackage[T1]{fontenc} 

\usepackage{filecontents} 

\usepackage{xr}

\newcommand*{\citen}{}
\DeclareRobustCommand*{\citen}[1]{%
  \begingroup
    \romannumeral-`\x 
    \setcitestyle{numbers}%
    \cite{#1}%
  \endgroup
}

\newcolumntype{.}{D{.}{.}{-1}}
\newcommand{\mc}[3]{\multicolumn{#1}{#2}{#3}}

\newcommand{\fns}{\footnotesize}

\newcommand{\fig}[2]{\scalebox{#1}{\includegraphics{#2}}}

\usepackage{bbm}

\usepackage{hyperref}
\hypersetup{colorlinks,breaklinks,linkcolor=Blue,urlcolor=Blue,anchorcolor=Blue,citecolor=Blue}
\usepackage{color}
\definecolor{DarkBlue}{rgb}{0.0,0.08,0.45}
\definecolor{Blue}{rgb}{0.0,0.0,1.0}
\definecolor{Red}{rgb}{1.0,0.0,0.0}
\definecolor{RedOrange}{rgb}{0.9,0.0,0.2}
\definecolor{dgrn}{RGB}{0,150,0}
\definecolor{dgray}{gray}{0.3}

\makeatletter
\newcommand*{\addFileDependency}[1]{
  \typeout{(#1)}
  \@addtofilelist{#1}
  \IfFileExists{#1}{}{\typeout{No file #1.}}
}
\makeatother

\newcommand*{\myexternaldocument}[1]{%
    \externaldocument{#1}%
    \addFileDependency{#1.tex}%
    \addFileDependency{#1.aux}%
}

\myexternaldocument{SI}

\usepackage{titlesec}  
\titleformat{\section}[hang]{\bfseries\large}{\thesection}{1em}{}
\renewcommand{\thesection}{\arabic{section}}
\renewcommand{\thesubsection}{.\arabic{subsection}}
\renewcommand{\thesubsubsection}{.\arabic{subsubsection}}
\titleformat{\section}{\normalfont\bfseries}{\thesection.~}{0.25em}{}
\titleformat{\subsection}[runin]{\normalfont\bfseries}{\thesection\thesubsection.}{0.25em}{}
\titleformat{\subsubsection}[runin]{\normalfont\bfseries}{\thesection\thesubsection\thesubsubsection.}{0.25em}{}
\begin{document}
\title{The Dose Makes the Poison: Perturbative Steps Toward the Ultimate Linearized Coupled Cluster Method}
\author{Sylvia J. Bintrim}
\author{Ella R. Ransford}
\author{Kevin Carter-Fenk*}
\affiliation{Department of Chemistry, University of Pittsburgh, Pittsburgh, Pennsylvania 15218, USA
\\
*\texttt{kay.carter-fenk@pitt.edu}
}
\date{\today}
\begin{abstract}
\begin{center}
\textbf{Abstract}
\end{center}
``Addition-by-subtraction’’ coupled cluster (CC) approaches provide a promising approach to treating the difficult strong correlation problem by simplifying the standard CC equations.
In a separate vein, linearized CC methods have drawn interest for their lower computational cost, increased parallelizability, and favorable properties for extension to the excited state--but the inclusion of ring/crossed-ring terms causes singularities even for single bond breaking. A linearized, addition-by-subtraction CC method called linearized ladder CCD (linLCCD) removes these terms to avoid divergences, but linLCCD under-estimates dynamical correlation. Herein we resolve this deficiency of linLCCD by introducing a linearized external coupled cluster perturbation theory
that adds a second-order ring/crossed-ring correction back into a linLCCD reference wave function. Our resultant xlinCCD(2) method is regular and yields comparable results to linearized CCD in weakly-correlated regimes.
\end{abstract}
\maketitle
\thispagestyle{plain}
\section*{Introduction}
\noindent
Strongly correlated systems ({\em e.g.} transition metal complexes or molecules undergoing bond dissociation)
are exceedingly difficult to accurately and affordably simulate using quantum chemistry methods. In such systems, no single electron configuration dominates the wave function; instead, the wave function is comprised of multiple (or many) nearly-degenerate configurations of roughly equal weights.
Full configuration interaction (FCI) offers the most straightforward solution to the strong correlation problem
by expanding the wave function in the complete basis of all Slater determinants, a procedure which
is exact but scales exponentially with system size. Due to its high cost, practitioners are forced into the
non-trivial selection of an ``active space'' of important orbitals for FCI-based methods.\cite{DasWah66,RooTaySig80,Roo80,SzaMulGid12}
While there are notable efforts to automate the selection of such orbitals,\cite{HuCha15, SaySunCha17, RenPenZha17, BaiRei20} the choice of active space
remains a source of uncontrolled error.\cite{ZhoGagTru2019}
In an effort to simultaneously avoid active space selection and reduce computational cost,
our group has been pursuing novel, single-reference coupled cluster (CC)  methods that capture the qualitative essence of strong correlation at polynomial cost.
Furthermore, we have shown that improvements to the CC ground state translate to improvements in the excited states of strongly correlated systems.\cite{BinCar25}

One seemingly paradoxical line of modern inquiry into treating the strong correlation problem in single-reference CC is
the simplification of the CC equations by removal of problematic components of the wave function.\cite{Bar24}
Such simplifications lead to a family of approximations known as addition-by-subtraction CC.
Examples of such approaches include the distinguishable cluster approximation, in which
exchange couplings between doubles clusters are neglected, allowing for
smooth dissociation of dinitrogen.\cite{KatMan13, RisPerBar16, RisPerBar19}
There has also been a recent surge of interest in seniority-zero
CC approaches such as the pair coupled cluster doubles (pCCD)\cite{SteHenScu14,HenBulSte14,BrzBogTec19,Bog21,Bar24,ChadeMBog24, JohFecNad25}
approximation, in which only paired double substitutions contribute to the CC wave function, yielding highly affordable,
single-reference methods that are robust in cases of static correlation.\cite{LimAyeJoh13,BogTecLim14,BogTecBul14,LimKimAye14, LesMatLeg22}
Despite the formal $\mathcal{O}(N^3)$ scaling of pCCD
(or equivalently, antisymmetric product of 1-reference orbital geminals),
it is not invariant to unitary transformations within the occupied or virtual orbitals,
requiring orbital optimization and localization to achieve size-consistent results.\cite{BogTecAye14, TecBogJoh14,BogTec17, NowTecBog19, BrzBogTec19} 

Singlet-paired and triplet-paired CCD (CCD0 or CCD1, respectively) methods
decouple the singlet- and triplet-paired doubles amplitudes
in efforts to attain similar reliability to pCCD while maintaining orbital invariance.\cite{BulHenScu15}
Perturbative recouplings (CCD with frozen singlet- or triplet-paired amplitudes [CCDf0/CCDf1]) can be introduced by fully optimizing one set of amplitudes
in the presence of the frozen amplitudes of the other.\cite{GomHenScu16, BinCar25} 
However, it remains unclear why the decoupling of singlet- and triplet-paired amplitudes in CCD0/CCD1
(and hence CCDf0/CCDf1) helps to avoid the failures of CCD and CCSD in strongly correlated systems.

In contrast, simplifications to the CC equations that use diagrammatic arguments to precisely
target terms for removal have a clear physical significance.\cite{ScuHenBul13, BulHenScu15}
One example is ring-CCD (or the particle-hole random phase approximation), which removes terms associated with
ladder diagrams and typically also exchange interactions to achieve somewhat better dissociation limits for chemical bonds.\cite{FucNiqGon05, VanYanYan13, TahRen19, For22}
Ladder-CCD or, equivalently, the particle-particle random phase approximation, 
is known to perform well for the low-density uniform electron gas (where electron-electron interactions are poorly-screened and thus strong), providing 
some physical explanation for its success in strongly correlated systems.\cite{YanvanYan13, YanvanSte13, SheAggYan14, VanYanYan14, YanPenLu14, TahRen19, LiJinYu24a, LiJinYu24b, MarRomLoo24, YuLiZhu25}

Recently, one of us applied the philosophy of addition-by-subtraction CC in conjunction with diagrammatic arguments to
improve the robustness of linearized CCD (linCCD) in strongly correlated cases.
While linCCD itself does not fall under the
addition-by-subtraction umbrella, it offers several advantages, including a straightforward variational framework,
simpler derivatives,
and improved parallel efficiency.\cite{Tau08, TauBar09, BarMusLot10} Despite these advantages, linCCD displays catastrophic divergences in strongly correlated systems.\cite{JanPal80}
While prior work involved regularizing the linCCD equations to suppress small energy-gap denominators,\cite{TauBar09}
our recent investigations instead suggest that ring and crossed-ring terms are to blame.\cite{Car25}
Removing the offending diagrams results in an addition-by-subtraction theory called
linearized ladder CCD (linLCCD)\cite{Car25} which is robust for strongly correlated systems
and has the favorable properties of unitary invariance and size consistency.
Despite the qualitative robustness of linLCCD in strongly correlated systems, it
lacks quantitative accuracy, missing particle-hole screening typically supplied by
ring and crossed-ring terms.

In this work, we improve upon linLCCD with a  linearized external coupled
cluster perturbation theory (xCCPT)\cite{LotBar11} correction
to reintroduce this missing correlation energy.
The resultant approach, which we call second-order external linearized CCD [xlinCCD(2)], incorporates all forms of linCCD correlation (driver, ladder, ring, and crossed-ring terms), and performs well for strongly correlated systems without sacrificing dynamical correlation.
Importantly, we choose a partitioning of the Hamiltonian that dresses the electron-repulsion integrals and one-particle energies
with correlation from the reference linLCCD wave function, stabilizing the addition of the ring and crossed-ring
terms that typically cause the divergence of infinite-order linCCD. In fact, our results feature
cases, such as the dissociation curve of dinitrogen, where the perturbative ring/crossed-ring terms and dressed one-particle energies actually \textit{prevent} divergence of the parent linLCCD theory.
As our xlinCCD(2) method contains all types of correlation present in linCCD but often
produces results of comparable accuracy to CCD in cases where linCCD diverges, we 
believe xlinCCD(2) is perhaps the most complete linearized coupled cluster doubles theory to date.
Given the widespread use of linCCD\cite{ZobSzaSur13, BogAye15, BogTec17, NowLegBog21, LesMatLeg22, ChaBogTec23, ChadeMBog24} or configuration interaction doubles\cite{NowBog23} in combination with pCCD, as well as multi-reference linCCD,\cite{ShaAla15, YiChe18, WaiSucKoh25} 
we expect that our results may encourage similar applications of xlinCCD(2).

\section*{Theoretical Background}
\noindent
Throughout this work, occupied
orbitals 
will be indexed as $\{i,j,k,l,\dots\}$,  virtual orbitals 
as $\{a,b,c,d,\dots\}$, and general indices as $\{m,n,p,q,\dots\}$.

The standard CCD approach employs an exponential
{\em ansatz} for the wave function,
\begin{equation}
|\Psi_{\text{CC}}\rangle = e^{\hat{T}_2}|\Phi_0\rangle\;~
\end{equation}
where $|\Phi_0\rangle$ is usually
the Hartree-Fock ground state reference determinant,
\begin{equation}
    \hat{T}_2 = \frac{1}{4}\sum\limits_{ijab}t_{ij}^{ab}\hat{a}_a^\dagger\hat{a}_b^\dagger\hat{a}_j\hat{a}_i
\end{equation}
is the double-substitution operator,
and $\hat{a}_i$ and $\hat{a}_a^\dagger$ are
particle annihilation and particle creation
operators, respectively.
The energy and amplitude equations
for CCD are
\begin{subequations}\label{eqn:main}
\begin{alignat}{1}
E = \langle\Phi_0|e^{-\hat{T}_2}\hat{H}e^{\hat{T}_2}|\Phi_0\rangle  &= \frac{1}{4}\sum_{ijab}t_{ij}^{ab}\bra{ij}\ket{ab}\label{subeqn:a}\\
\bra{\Phi_{ij}^{ab}}e^{-\hat{T}_2}\hat{H}e^{\hat{T}_2}\ket{\Phi_0}&=0
\label{subeqn:b}
\end{alignat}
\end{subequations}
where $\ket{\Phi_{ij}^{ab}}$ is a doubly excited determinant.

In linearized CCD (linCCD), we truncate the {\em ansatz} at first order in the Taylor expansion of $e^{\hat{T}_2}$, giving
\begin{equation}
|\Psi_{\text{CC}}\rangle = (1+\hat{T}_2)|\Phi_0\rangle\;~
\end{equation}
Whereas variational CC methods generally lead to non-terminating series,\cite{vanGor00, TauBar09}
we note that the linCCD energy functional can be written in Hermitian form 
\begin{equation}
\mathcal{E}=\bra{0}\lbrack (1+\hat{T}_2^\dagger)\hat{H}(1+\hat{T}_2)\rbrack_{\text{SC}}\ket{0}
\end{equation}
where ``SC'' denotes strongly connected diagrams.\cite{Tau08, TauBar09} 
Varying this expression with respect to $\hat{T}_2^\dagger$ leads to the following
doubles amplitude equation for linCCD at stationarity in the spin-orbital basis:
\begin{equation}
\begin{aligned}
0= & v_{i j}^{a b}-\mathcal{P}_{i j}\left(t_{k j}^{a b} f_i^k\right)+\mathcal{P}_{a b}\left(f_c^a t_{i j}^{c b}\right) \\
& +\frac{1}{2} t_{k l}^{a b} v_{i j}^{k l}+\frac{1}{2} v_{c d}^{a b} t_{i j}^{c d} \\
& +\mathcal{P}_{i j} \mathcal{P}_{a b}\left(v_{i c}^{a k} t_{k j}^{c b}\right)
\end{aligned}
\label{t2:CCD}
\end{equation}
where we have employed the Einstein summation convention,
$v_{pq}^{rs}$ are anti-symmetrized two-electron integrals $\bra{rs}\ket{pq}$ and
$\mathcal{P}_{pq}=1-(p\leftrightarrow q)$
are index permutation operators.
The connection of each term to a class of Feynman diagrams is as follows:
The first three
terms are known as ``driver'' terms, terms four and five
correspond to ``ladder'' diagrams,
and the final term encompasses ``ring'' and ``crossed-ring''
diagrams. Thus, omitting the final term leads to the linLCCD equations,
omitting the final term and the particle-particle ladder term (term 5) gives the so-called
hole-hole approximation to linLCCD [linLCCD(hh)], and eliminating everything
but the driver terms yields the second-order M{\o}ller-Plesset perturbation theory (MP2) equation.
For a detailed analysis of each term in the linCCD equations along with the corresponding diagrams, we refer the interested reader to
Ref.~\citen{Car25}, and for an overview of each diagram in the CCD equations we suggest Ref.~\citen{MasHumGru24}.

xCCPT perturbatively includes missing components of the (full) cluster operator $\hat{T}$ on top of an initial CC calculation that uses a (potentially incomplete) ``external'' $\hat{T}_{\text{X}}$ cluster operator.\cite{LotBar11}
Here, we introduce the linearization of the xCCPT equations for the first time.
Let $\hat{T}_{\text{X}}$ correspond to a linearized CCD starting point such as linLCCD.
We can partition the Hamiltonian into a one-electron part $\hat{H}_0$ and fluctuation potential $\hat{V}$:
\begin{equation}
    \hat{H}=\hat{H}_0+\lambda\hat{V}
\end{equation}
and define
\begin{equation}
\hat{T}=\hat{T}_{\text{X}}+\delta\hat{T}=\hat{T}_{\text{X}}+\sum_k\lambda^k\delta\hat{T}^{(k)}
\end{equation}
where $k$ denotes the order of the xCCPT correction to the wave function.
We choose 
\begin{equation}
\begin{split}
\hat{H}_0 &= \sum_{ik} \Big( f_i^k a_i^\dagger a_k + \frac{1}{2}\sum_{jab}t_{\text{X}ij}^{\phantom{\text{X}}ab}\bra{kj}\ket{ab}a_i^\dagger a_k\Big)\\
&\qquad +\sum_{ac} \Big(f_c^a a_a^\dagger a_c - \frac{1}{2}\sum_{ijb}t_{\text{X}ij}^{\phantom{\text{X}}cb}\bra{ij}\ket{ab} a_a^\dagger a_c\Big)
\end{split}
\label{eq:H0_mosaic}
\end{equation}
and modify $\hat{V}= \hat{H}-\hat{H}_0$ accordingly.
This choice of Hamiltonian partitioning 
dresses the one-particle energies with correlation from the reference wave function ({\em e.g.} linLCCD),
which has an important effect on the stability of the resultant perturbation theory (see Figure~\ref{fig:H0_choice}).

We will also let $\hat{X}=\hat{X}_0+\hat{X}_{\text{V}}$ where
\begin{equation}
    \hat{X}_0 = (1-\hat{T}_{\text{X}})\hat{H}_0(1+\hat{T}_{\text{X}})\approx \hat{H}_0+\lbrack \hat{H}_0, \hat{T}_X\rbrack
\end{equation}
and 
\begin{equation}
    \hat{X}_{\text{V}} = (1-\hat{T}_{\text{X}})\lambda\hat{V}(1+\hat{T}_{\text{X}})\approx \lambda \hat{V}+\lbrack \lambda \hat{V}, \hat{T}_X\rbrack
\end{equation}
are linearized, similarity-transformed $\hat{H}_0$ and $\hat{V}$ operators.
Having chosen our Hamiltonian partitioning and reference wave function,
we now derive the first order perturbative correction to the wave function
and second order correction to the energy via xCCPT.

We begin by writing the linCCD equation for the first order correction to the doubles amplitudes  $\delta \hat{T}:=\lambda \delta\hat{T}^{(1)}$: 
\begin{equation}
    \bra{\Phi_{ij}^{ab}}(1-\lambda\delta\hat{T})\hat{X}(1+\lambda\delta\hat{T})\ket{0}\approx \bra{\Phi_{ij}^{ab}}(\hat{X}+\lbrack \hat{X}, \lambda \delta \hat{T}\rbrack )\ket{0}=0
\end{equation}
This simplifies to
\begin{equation}
\begin{split}
\bra{\Phi_{ij}^{ab}}(\lbrack \hat{H}_0, \hat{T}_{\text{X}}\rbrack+\lambda\hat{V}+\lbrack \lambda\hat{V}, \hat{T}_{\text{X}}\rbrack)\ket{0}
+
\bra{\Phi_{ij}^{ab}}\lbrack\hat{H}_0, \lambda\delta\hat{T}\rbrack\ket{0}
=
0
\label{xlinccd2_linlccd_comm}
\end{split}
\end{equation}
when retaining only nonzero terms and truncating the perturbation series at $\lambda^1$.
After setting $\lambda=1$, Eq.~\ref{xlinccd2_linlccd_comm} implies 
\begin{equation}
\label{eq:15}
\begin{split}
0&=X_{ij}^{ab}+\mathcal{P}_{ab}(X_e^b\delta t_{ij}^{ae})-\mathcal{P}_{ij}(X_j^m\delta t_{im}^{ab})
\end{split}
\end{equation}
where 
\begin{equation}
\begin{split}
X_{ij}^{ab}&= v_{ij}^{ab} +\frac{1}{2}t_{\text{X}mn}^{\phantom{\text{X}}ab}v^{mn}_{ij}+\frac{1}{2}t_{\text{X}ij}^{\phantom{\text{X}}ef}v^{ab}_{ef}+\mathcal{P}_{ij}\mathcal{P}_{ab}(t_{\text{X}im}^{\phantom{\text{X}}ae}v^{mb}_{ej})
\\&+
\mathcal{P}_{ab}(f_e^bt_{\text{X}ij}^{\phantom{\text{X}}ae})-\mathcal{P}_{ij}(f_j^mt_{\text{X}im}^{\phantom{\text{X}}ab})
\end{split}
\label{Xijab}
\end{equation}
can be interpreted as a set of electron repulsion integrals that are screened by correlation effects from the reference wave function, and
\begin{subequations}
\begin{equation}
X_b^a=f_b^a-\frac{1}{2}t_{\text{X}mn}^{\phantom{\text{X}}ae}v^{mn}_{be}
\end{equation}
\begin{equation}
X_i^j=f_i^j+\frac{1}{2}t_{\text{X}im}^{\phantom{\text{X}}ef}v^{jm}_{ef}
\end{equation}
\label{eq:dressed_1p}
\end{subequations}
are dressed one-particle energies resulting from our choice of $\hat{H}_0$ that bear a resemblance to correlated orbital energies used in other approaches.\cite{ScuHenBul13,JahAhmBog25,CarHea23}
If $\hat{T}_{\text{X}}$ comes from a converged linLCCD calculation, then
\begin{equation}
0=v_{ij}^{ab} +\frac{1}{2}t_{\text{X}mn}^{\phantom{\text{X}}ab}v^{mn}_{ij}+\frac{1}{2}t_{\text{X}ij}^{\phantom{\text{X}}ef}v^{ab}_{ef}+
\mathcal{P}_{ab}(f_e^bt_{\text{X}ij}^{\phantom{\text{X}}ae})-\mathcal{P}_{ij}(f_j^mt_{\text{X}im}^{\phantom{\text{X}}ab})
\end{equation}
holds, and the first order amplitude correction equation (Eq.~\ref{eq:15}) simplifies to
\begin{equation}
0=\mathcal{P}_{ij}\mathcal{P}_{ab}(t_{\text{X}im}^{\phantom{\text{X}}ae}v^{mb}_{ej})+\mathcal{P}_{ab}(X_e^b\delta t_{ij}^{ae})
-\mathcal{P}_{ij}(X_j^m\delta t_{im}^{ab})
\label{eq:xlinccd2linlccd}
\end{equation}
where term 1 reintroduces the heretofore missing ring/crossed-ring correlation
and the last two, mosaic-style/disconnected terms\cite{ScuHenBul13, SheHenScu14} couple ring/crossed-ring correlation to ladder and driver components.
Note that in the limit that $T_{\text{X}}$ comes from linCCD, $X_{ij}^{ab}=0$ and so $\delta \hat{T} =0$,
as expected.

For xlinCCD(2), the second order energy correction is simply
\begin{equation}
\delta E^{(2)} = \frac{1}{4}\sum_{ijab}\bra{ij}\ket{ab}\delta t^{ab}_{ij}
\end{equation}
and to first order, the wavefunction is 
\begin{equation}
\ket{\Psi}\approx \ket{\Psi^{(0)}}+\ket{\Psi^{(1)}}=(1+\hat{T}_{\text{X}})\ket{0}+\delta\hat{T}\ket{0}=(1+\hat{T}_{\text{X}}+\delta\hat{T})\ket{0}
\end{equation}
\section*{Computational Details}
\noindent
All calculations reported here use locally modified versions of Q-Chem~v6.2\cite{QCHEM5} or the PySCF software package.\cite{PySCF1,PySCF2}
Dissociation curves
were computed in the aug-cc-pVTZ basis\cite{Dun89, KenDunHar92, WooDun93} 
and make use of both packages,
while Hubbard model calculations were performed in PySCF.
W4-11 calculations were carried out in Q-Chem.

Geometries and benchmark thermochemical energies are taken from the non-multi-reference subset of the W4-11 thermochemical database.\cite{W4-11}
To reduce errors from spin-contamination in the energies, we used self-consistently converged restricted open-shell Hartree-Fock (ROHF)
orbitals to build the unrestricted Fock matrix for input into the unrestricted CC equations.\cite{KnoAndAmo91}
To extrapolate our results to the complete basis set (CBS) limit, we use a two-point $n^{-3}$ extrapolation of the correlation energies with $n=3, 4$ for the aug-cc-pVnZ basis sets.\cite{HelKloKoc97}

Ozone vibrational frequencies were calculated using MP and CC methods in Q-Chem in the aug-cc-pVDZ basis set. The geometries and vibrational frequencies for linCCD, CCD, MP3, and xlinCCD(2) approaches were computed via finite difference whereas CCSD analytic gradients were used. The SCF and CC amplitude convergence tolerances were set to $10^{-10}$.

Singlet-triplet (S-T) gaps of transition-metal diatomics were calculated via $\Delta$MP and $\Delta$CC methods in Q-Chem in the def2-QZVPPD\cite{def2,def2aug} basis (no frozen core approximation).
Herein, we report TinySpins25: A set of theoretical best-estimates for S-T
gaps of 25 heteronuclear diatomic first- and second-row transition metal complexes.
Details on the composition of TinySpins25 can be found in the Supporting Information.
In brief, the S-T gaps in TinySpins25 were calculated using the MRCC\cite{MRCC} software package
at the CCSDT(Q)$_\Lambda$ level
with two-point extrapolation to the complete basis set limit.
As validation of this choice, which was motivated by the findings of Ref.~\citen{HaiTubLev19}, we
compared 7 complexes computed at the def2-TZVPPD level to the triple-$\zeta$
FCI best estimates in the Quest~\#8 data set and find a mean absolute error of just 0.05~eV.

Li$^+$/ethylene carbonate (EC) cluster association energies $\Delta U_{\rm assoc}=U_{\text{Li}_x\text{EC}_y}-xU_{\text{Li}^+}-yU_{\text{EC}}$ were calculated in the def2-TZVPD basis with the resolution of the identity, using Q-Chem. For CC calculations with \#EC$\geq 3$, a two- or three-point extrapolation with FNOs was used with natural orbital occupation thresholds of 99.50\%, 99.75\%, and 99.80\%.\cite{LanKhiDol10, PokIzmKry10} By comparison to canonical CC results, the 99.50\%/99.75\% occupation threshold extrapolation errors were 2 kcal/mol and 5 kcal/mol for LiEC and LiEC$_3$, respectively.
Given that the benchmark DLPNO-CCSD(T) data also may contain substantial localization error, we believe that these extrapolation errors are sufficiently small.\cite{GraHer24}
Cluster geometries optimized using $\omega$B97X-D3BJ/def2-TZVPD and DLPNO-CCSD(T)/aug-cc-pVTZ association energy data were taken from Ref.~\citen{SteAgaBha25}. 

For the BDEs of first-row transition metal diatomics, we used ROHF with the exact-two-component (X2C) scalar relativistic approximation\cite{IliSau07, LiuPen09, Sau11, LiXiaLiu12} for input into unrestricted CC, extrapolating the SCF\cite{HalHelJor99} and correlation energies\cite{Tru98, NeeVal11} using def2-TZVPP and def2-QZVPP\cite{Wei06} basis sets. Spatial symmetry was not utilized at the SCF or CC steps. Reference bond lengths, spin states, and BDEs were taken from Refs.~\citen{FurPer06} and \citen{Abdul}.  
Spin-orbit coupling corrections from Ref.~\citen{Abdul} were applied.
\section*{Results \& Discussion}
\begin{figure}[ht!]
\centering
\fig{1.0}{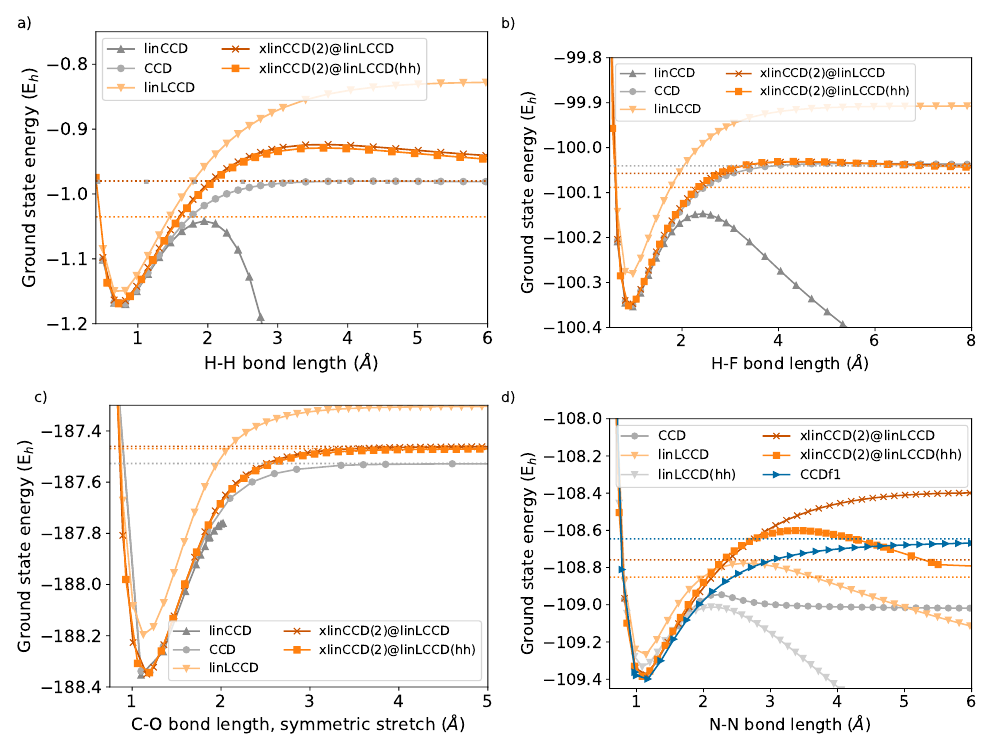}
\caption{
Ground state dissociation curves of
    (a) H$_2$, (b) FH, (c) CO$_2$ undergoing symmetric stretch, and (d) spatially symmetry-adapted N$_2$ in the aug-cc-pVTZ basis. The dotted horizontal lines indicate asymptotic limits estimated at 100 \r{A},
    except for the case of CCDf1, where the limit was estimated at 12~\r{A}.}\label{fig:xccpt}
\end{figure}
\noindent
Covalent bond breaking is a classic case where absolute near-degeneracy static correlation is encountered,\cite{HolGil11}
resulting in neccessarily multi-reference character of the wave function.
Here we analyze the performance of our single-reference xlinCCD(2) method in producing qualitatively correct dissociation curves for several small molecules.

We begin with H$_2$ in Figure~\ref{fig:xccpt}a, where linCCD diverges after 2~\AA\ separation. While linLCCD produces a smooth dissociation curve, it is significantly under-correlated by comparison to CCD, even at the equilibrium geometry. We show dissociation curves from two varieties of xlinCCD(2): one built atop a linLCCD reference that we call xlinCCD(2)@linLCCD, and another that uses a linLCCD(hh) reference, called xlinCCD(2)@linLCCD(hh). The latter is potentially more affordable with a simple change of basis (See Supporting Information for details on its computational cost.). Compared with linLCCD, both xlinCCD(2) methods provide an equilibrium energy and a finite asymptotic dissociation energy closer to those of CCD, albeit with a dissociation barrier. The asymptotic limit of xlinCCD(2)@linLCCD is a near-perfect match to CCD, differing by only 0.5~kcal/mol. This small imperfection is absent in the minimal-basis dissociation curve shown in Figure~\ref{fig:H2minbasis}, as xlinCCD(2)@linLCCD is exact in this limit.
The results for the hetero-diatomic FH dissociation curve in Figure~\ref{fig:xccpt}b are qualitatively similar to H$_2$.

For the symmetric dissociation of CO$_2$, the results in Figure~\ref{fig:xccpt}c suggest that the xlinCCD(2)
methods find a slightly higher dissociation limit than CCD, which may be attributable to the lack of quadratic terms (disconnected quadruples) that are important in the dissociation of double bonds.
While both are an upper bound to CCD,
xlinCCD(2)@linLCCD(hh) 
gives an energy at dissociation that is slightly closer to the CCD reference. 

For the dissociation of the N$_2$  triple bond (Figure~\ref{fig:xccpt}d), no theory that truncates at doubles can be expected to provide quantitative accuracy.\cite{KrySheByr98,MusBar05} Our previous addition-by-subtraction CC method of choice,\cite{BinCar25} CCDf1, lacks the unphysical barrier produced by xlinCCD(2) or CCD. However, compared with CCDf1, xlinCCD(2) provides dissociation energies that are closer to the CCD result. Interestingly, xlinCCD(2) does not diverge, even though the underlying linLCCD and linLCCD(hh) both dive downward for this triple bond. Of the two xlinCCD(2) methods, xlinCCD(2)@linLCCD(hh) appears to give a better asymptotic energy, even though linLCCD(hh) plunges downward more severely than linLCCD.
Notably, very few (if any) methods that are linear in the wave function can dissociate spatial symmetry-adapted N$_2$
without resorting to some form of regularization.

\begin{figure}[ht!]
    \centering
    \fig{1.0}{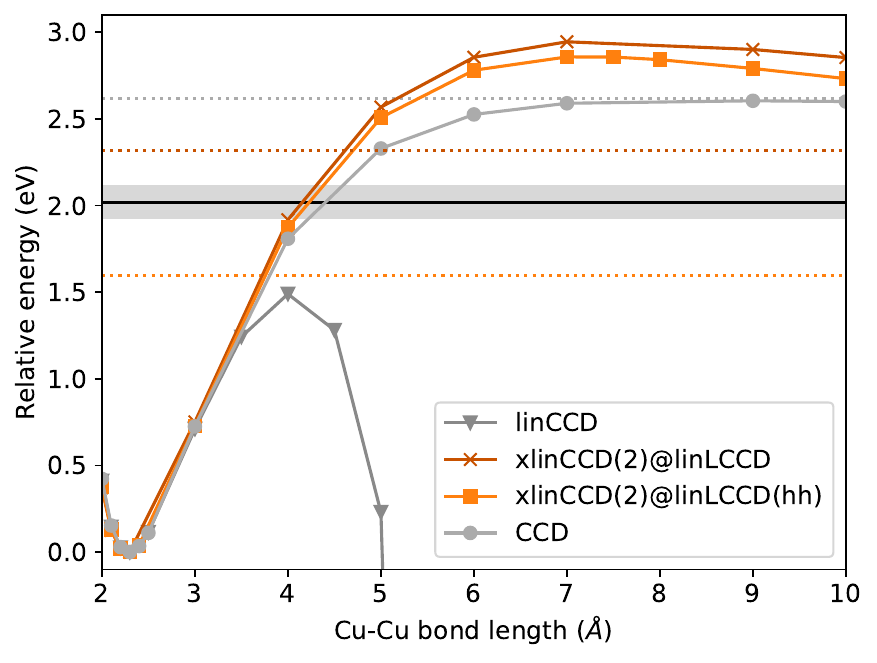}
\caption{Ground state dissociation curves for Cu$_2$ in the aug-cc-pVTZ basis. The dotted horizontal lines indicate the methods' asymptotic limits calculated at $100$ \r{A}. The experimental dissociation energy is shown as a black line, with experimental uncertainty as a gray region.\cite{DroHon57}
    }
\label{fig:xccpt_Cu}
\end{figure}
Even for a transition metal diatomic like Cu$_2$ (Figure~\ref{fig:xccpt_Cu}), xlinCCD(2) avoids divergence in the dissociation limit and tracks the CCD curve reasonably well. 
Whereas linCCD diverges, xlinCCD(2) not only converges to a clear asymptotic limit but outperforms CCD in estimating the bond dissociation energy (BDE) relative to experiment.

Our data for the Cu$_2$ BDE inspired us to more thoroughly assess the performance of xlinCCD(2) for BDEs and other thermochemical data in the  non-multireference subset of the W4-11 database.\cite{W4-11} While these explicitly non-multireference thermochemical results do not directly probe the performance of xlinCCD(2) for strong correlation, our findings (Figure~\ref{fig:w411}) reinforce that xlinCCD(2) consistently provides results of a quality comparable to those of linCCD or CCD at equilibrium geometries and less strongly correlated systems.

To assess the quality of potential energy surface shape provided by our methods within the Franck-Condon region, we computed frequencies of the three vibrational modes of ozone. The ground state ozone vibrational asymmetric stretch (highest-frequency mode) is known to be especially computationally challenging due to static correlation.\cite{StaLipMag89, LeeScu90} For the ozone asymmetric stretch, Table~\ref{tab:ozone_vib} shows that the xlinCCD(2) methods perform noticeably better than linCCD and are on par with estimates from MP3 and CCD. The $t_1$ amplitudes seem to play an important role in this vibrational mode, as we see much improvement in going from 
CCD to CCSD. This motivates further extension of xlinCCD(2) to include singles amplitudes in future work.
\begin{table}[h]
\caption{Harmonic frequencies (cm$^{-1}$) for the vibrational modes of ozone$^a$}
\label{tab:ozone_vib}
\begin{tabular}{l|l|l|l|l|l|l|l|l|l}
\hline\hline
MP2 & MP3 & MP4$^b$ & \makecell{xlinCCD(2)@\\linLCCD(hh)} & \makecell{xlinCCD(2)@\\linLCCD} 
& linCCD & CCD & CCSD & exp.$^b$ \\
\hline
738
& 796 & 695 & 787 & 785 
& 750 
& 787
& 753
& 716 \\
\hline
1149
& 1349 & 1123 & 1317 & 1308 
& 1198 
& 1324
& 1218
& 1135 \\
\hline
\cellcolor{Red!30}\textbf{2260}
& \cellcolor{Yellow!30}\textbf{1723} 
& \cellcolor{LimeGreen!30}\textbf{1547} 
& \cellcolor{Yellow!30}\textbf{1740} 
& \cellcolor{Yellow!30}\textbf{1758} 
&\cellcolor{Orange!30}\textbf{1961} 
& \cellcolor{GreenYellow!30}\textbf{1686}
& 
\cellcolor{Green!30}\textbf{1248}
& \textbf{1089} \\
\hline\hline
\mc{6}{l}{\fns $^a$Our calculations were performed in the aug-cc-pVDZ basis.}
\\
\mc{6}{l}{\fns $^b$MP4 and experimental frequencies from Ref.~\citen{StaLipMag89}}
\end{tabular}
\end{table}
\begin{figure}[ht!]
    \centering
    \fig{1.0}{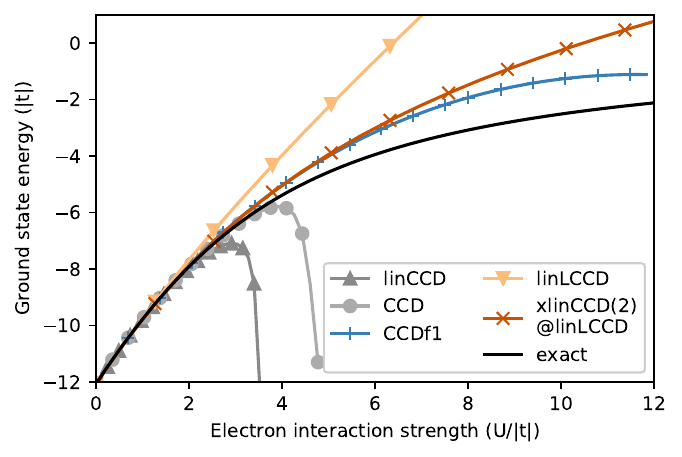}
\caption{Ground state energies
    as a function of interaction
    strength ($U/|t=-1.5|$) for a 10-site,
    half-filled Hubbard model
    with open boundary conditions.
    The exact reference is the FCI result.}
    \label{fig:hubbard}
\end{figure}

Next, we consider 
the Hubbard model,\cite{Hub63} allowing us to directly modulate the interaction strength between electrons at different sites
to assess the behavior of our methods in weak to strong correlation regimes.
In Figure~\ref{fig:hubbard}, we plot the ground state energy of a ten-site, one-dimensional, molecular Hubbard model at half-filling as a function of interaction strength (U/|t|).
Our goal here is to assess how well various single-reference CC methods can capture the physics of the strongly correlated Hubbard model
by tracking which methods can qualitatively reproduce the exact FCI result out to higher electron interaction strengths.

Figure~\ref{fig:hubbard} shows that the linCCD energy diverges around $U/|t|=3$ while CCD diverges slightly later, also monotonically decreasing with interaction strength, after  $U/|t|\sim 4$. LinLCCD does not diverge downward but appears almost as under-correlated as Hartree-Fock.\cite{BinCar25}
Although CCDf1 tracks the exact result more closely than xlinCCD(2)@linLCCD, the latter method does not ``turn over'' at high interaction strengths, making it potentially better-suited to serve as a ground state reference for excited state methods. Based on findings in our previous work, we suspect that if both ground and excited state methods avoid the turn-over, the resulting excitation energies will be closer to the exact result.\cite{BinCar25}

Inspired by the promising results from xlinCCD(2) applied to the Hubbard model, we calculated BDEs for first-row transition metal diatomics of various correlation strengths, including metal hydrides, chlorides, and oxides. 
For these molecules, both xlinCCD(2) methods performed on par with linCCD and CCD. Interestingly, linLCCD(hh) provided the overall smallest mean absolute error (Table~\ref{tab:tm_diatomics}). While 
K$_2$, Zn$_2$, and the closed-shell Ni$_2$ and Cu$_2$ can reasonably be treated with CCD and linCCD, qualitative accuracy for the remaining first-row transition metal homonuclear diatomics mandates $t_1$ amplitudes as well as perturbative triples amplitudes, at minimum.\cite{HaiTubLev19}
With that in mind, it is encouraging that for the heteronuclear transition metal diatomics, xlinCCD(2)@linLCCD outperforms linCCD by 0.9~kcal/mol, coming within 0.3~kcal/mol of CCSD.
\begin{figure}[ht!]
    \centering
\fig{1.0}{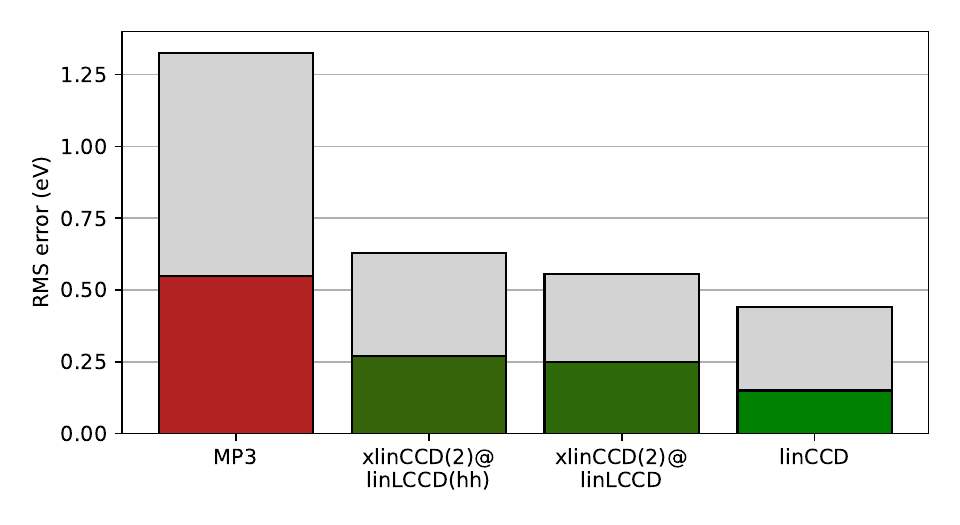}
\caption{Root-mean-square error (eV) (colored bars) and maximum errors (gray bars) for the lowest energy singlet/triplet gaps of 25 transition metal diatomics as computed by $\Delta$CC and $\Delta$MP2 methods in the def2-QZVPPD basis. 
}
\label{fig:STgaps}
\end{figure}
As a first foray into excited state properties with xlinCCD(2), we computed the first singlet-triplet (S-T) gaps via $\Delta$MP and $\Delta$CC methods for our new dataset (TinySpins25) of 25 transition metal diatomics containing metals Ag, Au, Cd, Cu, Hg, Pt, Ru, Sc, and Zn. The theoretical best estimate (TBE) gaps for TinySpins25 are provided in the Supporting Information. The gaps are fairly small (averaging to 1.35 eV and ranging from 0.09 eV to 3.38 eV), making this a challenging data set for single-reference methods even though the $T_1$-diagnostic\cite{LeeTay89} never exceeds 0.05.\cite{JiaDeYNat11} The xlinCCD(2) methods are theoretically similar in computational cost to MP3 but provide much smaller errors for the gaps (Figure~\ref{fig:STgaps}). For these equilibrium-geometry diatomics, linCCD provides excellent results within chemical accuracy. Even so, our bond dissociation curves ({\em e.g.} Figure~\ref{fig:xccpt_Cu}) suggest that the xlinCCD(2) methods will out-compete linCCD by providing robust results not just at the equilibrium geometry but also during bond dissociation. The xlinCCD(2) correction reduces the S-T gap root-mean-square error to 0.25 eV, down from 0.40 eV for linLCCD (see Supporting Information). Similarly, the error in the xlinCCD(2)@linLCCD(hh) gaps has been reduced to 0.27 eV, down from 0.41 eV for linLCCD(hh) (see Supporting Information), so we can conclude that making the more affordable hole-hole approximation at the linLCCD step does not have much of an effect on the accuracy of the gaps.

Solvated Li$^+$ clusters 
pose a surprisingly challenging problem for many quantum chemistry methods, providing an interesting, industrially relevant test system for our new xlinCCD(2) methods. Of the density functionals tested by Stevensen et al., the poor performance of B2PLYP-D3 (mean absolute error of 5.9~kcal/mol)
relative to lower-rung functionals like $\omega $B97X-D3(BJ) and CAM-B3LYP-D3
on the problem of predicting
association energies for Li$^+$/ethylene carbonate (EC) clusters caught our attention. We suspect that the failure of B2PLYP-D3 is due to
its dependence on the MP2 wave function, and our suspicion is corroborated by our finding that MP2 
provides a mean absolute error of 10.5 kcal/mol for this problem (Table~\ref{tab:LiEC}).

Overall, we find that xlinCCD(2) provides results within our expectations for Li$^+$/ethylene carbonate clusters.
The hole-hole approximation to the linLCCD reference wave function has little effect on the accuracy of the xlinCCD(2) results, which are of CCD-level quality (within 0.1~kcal/mol of each other). Both reference wave functions for xlinCCD(2) also lead to 1~kcal/mol improvements over standard linCCD.

While likely not quite as affordable as MP2 even if implemented with our one-shot, memory-efficient strategy,\cite{Car25} linLCCD(hh) (mean absolute error 5.5 kcal/mol) performs substantially better than MP2 (10.5 kcal/mol error), suggesting that a double hybrid functional based upon linLCCD(hh) might be capable of providing better results for these lithium clusters. Such a density functional is forthcoming from  our research group.\cite{RanCar26} In the immediate context, this result implies that linLCCD(hh) wave functions are a better jumping-off point for
higher-order perturbative corrections.

\begin{table}[h]
\caption{Energies of association of Li/EC clusters in kcal/mol}
\label{tab:LiEC}
\begin{tabular}{l l . . . . . . . . .}
\hline\hline
\mc{3}{c}{Benchmark} && \mc{7}{c}{Errors$^a$}\\ \cline{1-3} \cline{5-11}
\mc{1}{c}{\#Li}
&\mc{1}{c}{\#EC}
&\mc{1}{c}{CCSD(T)$^b$} 
&&\mc{1}{c}{MP2} 
&\mc{1}{c}{linLCCD(hh)}
&\mc{1}{c}{linLCCD}
&\mc{1}{c}{xlinCCD(2)$^c$}&\mc{1}{c}{xlinCCD(2)$^d$} 
&\mc{1}{c}{linCCD}
&\mc{1}{c}{CCD}\\
\hline
1 & 1 & -51.3 && 3.4 & 1.8 & 0.4 & 1.5 & 1.6 & 2.3 & 1.7\\
\hline
1 & 2 & -90.1 && 5.6 & 2.8 & 0.4 & 2.2 & 2.4 & 2.9 & 0.9\\
\hline
1 & 3 & -116.0 && 7.5 & 4.1 & 1.4 & 3.5 & 2.3 & 5.4 & 4.1\\
\hline
1 & 4 & -129.9 && 2.0 & 1.6 & 9.7 & 8.8 & 8.5 & 6.4 & 7.5\\
\hline
2 & 4 & -157.9 && 13.3 & 9.6 & 0.7 & 1.1 & 1.5 & 3.7 & 2.6\\
\hline
3 & 4 & -101.1 && 17.6 & 13.0 & 2.9 & 3.9 & 4.3 & 7.0 & 5.2\\
\hline
4 & 6 & -97.4 && 24.0 & 5.9\\
\hline
 & & \text{MAE} && \cellcolor{red!30}10.5 & \cellcolor{yellow!30}5.5 & \cellcolor{yellow!30}2.6 & \cellcolor{yellow!30}3.5 & \cellcolor{yellow!30}3.4 & \cellcolor{yellow!30}4.6 & \cellcolor{yellow!30}3.6
\\
\hline\hline
\mc{11}{l}{\fns $^a$All calculations done in the Def2-TZVPD basis set}\\
\mc{11}{l}{\fns $^b$DLPNO-CCSD(T)/aug-cc-pVTZ data from Ref.~\citen{SteAgaBha25}}\\
\mc{11}{l}{\fns $^c$linLCCD(hh) reference wave function}\\
\mc{11}{l}{\fns $^d$linLCCD reference wave function}
\end{tabular}
\end{table}
Finally, we have numerically verified the important property of size consistency for
xlinCCD(2)\allowbreak@linLCCD and xlinCCD(2)\allowbreak@linLCCD(hh) by considering the case of two hydrogen dimers. 
To see this analytically for the first method, suppose we simultaneously diagonalize $X_i^j$ and $X_b^a$ ($X_i^j \oplus X_b^a$)
so that Eq.~\ref{eq:xlinccd2linlccd} becomes
\begin{equation}
\delta t_{ij}^{ab}=-\frac{\mathcal{P}_{ij}\mathcal{P}_{ab}(\tilde{t}_{Xim}^{\:\:\:ae}\tilde{v}^{mb}_{ej})}{2\tilde{X}_b-2\tilde{X}_j}
\end{equation}
in the new basis $\tilde{X}_j \oplus \tilde{X}_b$ of eigenvectors of $X_i^j \oplus X_b^a$. Note that the block-diagonal basis transformation only mixes occupied orbitals with other occupied orbitals and likewise for virtual orbitals. 

As linLCCD is size consistent, for any disjoint $i\rightarrow a$ and $m\rightarrow e$ excitation pairs localized on well-separated fragments, $\tilde{t}_{Xim}^{\:\:\:ae}$ will be zero. Similarly, for any disjoint $j\rightarrow b$ and $m\rightarrow e$, $\tilde{v}_{ej}^{mb}$ will be zero. By transitivity, $\delta t_{ij}^{ab}$ is zero for any disjoint $i\rightarrow a$ and $j\rightarrow b$ excitation pairs. A similar argument holds for xlinCCD(2)@linLCCD(hh), since linLCCD(hh) is also size consistent.\cite{Car25} (See Supporting Information.) We note that xlinCCD(2) is a coupled electron pair theory that contributes no correlation between excitation pairs on disjoint molecular fragments.
\section*{Conclusion}
\noindent
In conclusion, we have presented a size-consistent, perturbative correction to linCCD called  xlinCCD(2). Our approach can take
any reference wave function as input, but we have tested the specific choices of linLCCD and linLCCD(hh).
Via calculations of thermochemical properties, bond dissociation energies and singlet-triplet gaps of transition metal diatomics, and
the ozone asymmetric stretch vibrational mode, we have shown that xlinCCD(2) provides results of quality comparable to CCD. We also find that xlinCCD(2) produces CCD-quality results for strongly correlated systems well beyond the equilibrium geometry. xlinCCD(2) is capable of dissociating covalent
bonds of homonuclear diatomics and performs well for the Hubbard model at high interaction strength. In general, xlinCCD(2) performs as well as linCCD at equilibrium but vastly outperforms it beyond the Condon region.

Next steps include algorithmic enhancements of the efficiency of xlinCCD(2) via one-shot implementations proposed in the Supporting Information.
Apart from appearing more amenable to tensor hypercontraction density fitting algorithms than the usual iterative CC approaches,\cite{Pie25}
such one-shot algorithms could also avoid numerical instabilities that plague nonlinear amplitude equations.\cite{SonGonYe25}
Given the clear importance of singles amplitudes in many of the chemical systems explored, we are currently 
formulating a related, $t_1$-inclusive xCCPT approach. Overall, our results suggest it is possible to rescue single-reference linCCD approaches for strongly correlated systems.
{\section*{Supporting Information}
The Supporting Information is available free of charge at ...
\begin{itemize}
\item[] Numerical PES data (XLSX)
\item[] W4-11 thermochemical data (XLSX)
\item[] BDEs, bond lengths, and spin states for first-row transition metal diatomics (XLSX)
\item[] S-T gaps and bond lengths for transition metal diatomics (TinySpins25) (XLSX)
\end{itemize}
\section*{Acknowledgements}
We thank Abdulrahman Y. Zamani for many enlightening discussions, assistance with Q-Chem input files, and guidance to helpful references. We also thank Shawna Sinchak for her proof-of-concept, memory-efficient, one-shot implementation of linLCCD within the hole-hole approximation. S.J.B. thanks Ethan Vo for suggesting 
calculations 
and Zachary K. Goldsmith for providing optimized Li$^+$/ethylene carbonate cluster geometries from Ref.~\citen{SteAgaBha25}. We thank Md. Rafi Ul Azam for providing feedback on this manuscript.
E. R. R. acknowledges support from the Wass Undergraduate Research Fellowship.
This research was supported in part by the University of Pittsburgh
and the University of Pittsburgh Center for Research Computing, RRID:SCR\_022735, through the resources provided.
Specifically, this work used the H2P cluster, which is supported by NSF award number OAC-2117681.

\newpage



\providecommand{\refin}[1]{\\ \textbf{Referenced in:} #1}
\begin{thebibliography}{100}

\bibitem{DasWah66}
Das,~G.;\ \ Wahl,~A.~C. Extended {Hartree-Fock} wavefunctions: {Optimized} valence configurations for h$_{\text{2}}$ and li$_{\text{2}}$, optimized double configurations for f$_{\text{2}}$. \textit{J.~Chem.\ Phys.} \textbf{1966,} \textsl{44,} 87--96.

\bibitem{RooTaySig80}
Roos,~B.~O.;\ \ Taylor,~P.~R.;\ \ Sigbahn,~P.~E. A complete active space {SCF} method ({CASSCF}) using a density matrix formulated super-{CI} approach. \textit{Chem.\ Phys.} \textbf{1980,} \textsl{48,} 157--173.

\bibitem{Roo80}
Roos,~B.~O. The complete active space {SCF} method in a fock-matrix-based super-{CI} formulation. \textit{Int.~J. Quantum Chem.} \textbf{1980,} \textsl{18,} 175--189.

\bibitem{SzaMulGid12}
Szalay,~P.~G.;\ \ M\"uller,~T.;\ \ Gidofalvi,~G.;\ \ Lischka,~H.;\ \ Shepard,~R. Multiconfiguration self-consistent field and multireference configuration interaction methods and applications. \textit{Chem.\ Rev.} \textbf{2012,} \textsl{112,} 108--181.

\bibitem{HuCha15}
Hu,~W.;\ \ Chan,~G. K.-L. Excited-state geometry optimization with the density matrix renormalization group, as applied to polyenes. \textit{J.~Chem.\ Theory Comput.} \textbf{2015,} \textsl{11,} 3000--3009.

\bibitem{SaySunCha17}
Sayfutyarova,~E.~R.;\ \ Sun,~Q.;\ \ Chan,~G. K.-L.;\ \ Knizia,~G. Automated construction of molecular active spaces from atomic valence orbitals. \textit{J.~Chem.\ Theory Comput.} \textbf{2017,} \textsl{13,} 4063--4078.

\bibitem{RenPenZha17}
Ren,~J.;\ \ Peng,~Q.;\ \ Zhang,~X.;\ \ Yi,~Y.;\ \ Shuai,~Z. Role of the dark 2ag state in donor--acceptor copolymers as a pathway for singlet fission: A dmrg study. \textit{J.~Phys.\ Chem.\ Lett.} \textbf{2017,} \textsl{8,} 2175--2181.

\bibitem{BaiRei20}
Baiardi,~A.;\ \ Reiher,~M. The density matrix renormalization group in chemistry and molecular physics: {Recent} developments and new challenges. \textit{J.~Chem.\ Phys.} \textbf{2020,} \textsl{152,} 040903:1--22.

\bibitem{ZhoGagTru2019}
Zhou,~C.;\ \ Gagliardi,~L.;\ \ Truhlar,~D.~G. Multiconfiguration pair-density functional theory for iron porphyrin with {CAS, RAS}, and {DMRG} active spaces. \textit{J.~Phys.\ Chem.~A} \textbf{2019,} \textsl{123,} 3389--3394.

\bibitem{BinCar25}
Bintrim,~S.~J.;\ \ {Carter}-{Fenk},~K. Optimal-reference excited state methods: {Static} correlation at polynomial cost with single-reference coupled-cluster approaches. \textit{J.~Chem.\ Theory Comput.} \textbf{2025,} \textsl{21,} 4080--4094.

\bibitem{Bar24}
Bartlett,~R.~J. Perspective on {Coupled}-cluster {Theory}. {The} evolution toward simplicity in quantum chemistry. \textit{Phys.\ Chem.\ Chem.\ Phys.} \textbf{2024,} \textsl{26,} 8013--8037.

\bibitem{KatMan13}
Kats,~D.;\ \ Manby,~F.~R. Communication: The distinguishable cluster approximation. \textit{J.~Chem.\ Phys.} \textbf{2013,} \textsl{139,} 021102:1--4.

\bibitem{RisPerBar16}
Rishi,~V.;\ \ Perera,~A.;\ \ Bartlett,~R.~J. Assessing the distinguishable cluster approximation based on the triple bond-breaking in the nitrogen molecule. \textit{J.~Chem.\ Phys.} \textbf{2016,} \textsl{144,}.

\bibitem{RisPerBar19}
Rishi,~V.;\ \ Perera,~A.;\ \ Bartlett,~R.~J. Behind the success of modified coupled-cluster methods: addition by subtraction. \textit{Molecular Physics} \textbf{2019,} \textsl{117,} 2201--2216.

\bibitem{SteHenScu14}
Stein,~T.;\ \ Henderson,~T.~M.;\ \ Scuseria,~G.~E. Seniority zero pair coupled cluster doubles theory. \textit{J.~Chem.\ Phys.} \textbf{2014,} \textsl{140,} 214113:1--8.

\bibitem{HenBulSte14}
Henderson,~T.~M.;\ \ Bulik,~I.~W.;\ \ Stein,~T.;\ \ Scuseria,~G.~E. Seniority-based coupled cluster theory. \textit{J.~Chem.\ Phys.} \textbf{2014,} \textsl{141,} 244104:1--10.

\bibitem{BrzBogTec19}
Brz{\c e}k,~F.;\ \ Boguslawski,~K.;\ \ Tecmer,~P.;\ \ {\.Z}uchowski,~P.~S. Benchmarking the accuracy of seniority-zero wave function methods for noncovalent interactions. \textit{J.~Chem.\ Theory Comput.} \textbf{2019,} \textsl{15,} 4021--4035.

\bibitem{Bog21}
Boguslawski,~K. Open-shell extensions to closed-shell {pCCD}. \textit{Chem.\ Commun.} \textbf{2021,} \textsl{57,} 12277--12280.

\bibitem{ChadeMBog24}
Chakraborty,~R.;\ \ de~Moraes,~M. M.~F.;\ \ Boguslawski,~K.;\ \ Nowak,~A.;\ \ Swierczynski,~J.;\ \ Tecmer,~P. Toward reliable dipole moments without single excitations: The role of orbital rotations and dynamical correlation. \textit{J.~Chem.\ Theory Comput.} \textbf{2024,} \textsl{20,} 4689--4702.

\bibitem{JohFecNad25}
Johnson,~P.~A.;\ \ Fecteau,~C.-{\'E}.;\ \ Nadeau,~S.;\ \ Rodr{\'\i}guez-Mayorga,~M.;\ \ Loos,~P.-F. Connections between {R}ichardson-{G}audin states, perfect-pairing, and pair coupled-cluster theory. \textit{arXiv preprint arXiv:2510.06144} \textbf{2025,} .

\bibitem{LimAyeJoh13}
Limacher,~P.~A.;\ \ Ayers,~P.~W.;\ \ Johnson,~P.~A.;\ \ De~Baerdemacker,~S.;\ \ Van~Neck,~D.;\ \ Bultinck,~P. A new mean-field method suitable for strongly correlated electrons: {Computationally} facile antisymmetric products of nonorthogonal geminals. \textit{J.~Chem.\ Theory Comput.} \textbf{2013,} \textsl{9,} 1394--1401.

\bibitem{BogTecLim14}
Boguslawski,~K.;\ \ Tecmer,~P.;\ \ Limacher,~P.~A.;\ \ Johnson,~P.~A.;\ \ Ayers,~P.~W.;\ \ Bultinck,~P.;\ \ De~Baerdemacker,~S.;\ \ Van~Neck,~D. Projected seniority-two orbital optimization of the antisymmetric product of one-reference orbital geminal. \textit{J.~Chem.\ Phys.} \textbf{2014,} \textsl{140,} 214114:1--8.

\bibitem{BogTecBul14}
Boguslawski,~K.;\ \ Tecmer,~P.;\ \ Bultinck,~P.;\ \ De~Baerdemacker,~S.;\ \ Van~Neck,~D.;\ \ Ayers,~P.~W. Nonvariational orbital optimization techniques for the {AP}1ro{G} wave function. \textit{J.~Chem.\ Theory Comput.} \textbf{2014,} \textsl{10,} 4873--4882.

\bibitem{LimKimAye14}
Limacher,~P.~A.;\ \ Kim,~T.~D.;\ \ Ayers,~P.~W.;\ \ Johnson,~P.~A.;\ \ De~Baerdemacker,~S.;\ \ Van~Neck,~D.;\ \ Bultinck,~P. The influence of orbital rotation on the energy of closed-shell wavefunctions. \textit{Mol.\ Phys.} \textbf{2014,} \textsl{112,} 853--862.

\bibitem{LesMatLeg22}
Leszczyk,~A.;\ \ M{\'a}t{\'e},~M.;\ \ Legeza,~{\"O}.;\ \ Boguslawski,~K. Assessing the accuracy of tailored coupled cluster methods corrected by electronic wave functions of polynomial cost. \textit{J.~Chem.\ Theory Comput.} \textbf{2022,} \textsl{18,} 96-117 PMID: 34965121.

\bibitem{BogTecAye14}
Boguslawski,~K.;\ \ Tecmer,~P.;\ \ Ayers,~P.~W.;\ \ Bultinck,~P.;\ \ De~Baerdemacker,~S.;\ \ Van~Neck,~D. Efficient description of strongly correlated electrons with mean-field cost. \textit{Phys.\ Rev.~B} \textbf{2014,} \textsl{89,} 201106:1--4.

\bibitem{TecBogJoh14}
Tecmer,~P.;\ \ Boguslawski,~K.;\ \ Johnson,~P.~A.;\ \ Limacher,~P.~A.;\ \ Chan,~M.;\ \ Verstraelen,~T.;\ \ Ayers,~P.~W. Assessing the accuracy of new geminal-based approaches. \textit{J.~Phys.\ Chem.~A} \textbf{2014,} \textsl{118,} 9058--9068.

\bibitem{BogTec17}
Boguslawski,~K.;\ \ Tecmer,~P. Benchmark of {Dynamic} {Electron} {Correlation} {Models} for {Seniority}-{Zero} {Wave} {Functions} and {Their} {Application} to {Thermochemistry}. \textit{J.~Chem.\ Theory Comput.} \textbf{2017,} \textsl{13,} 5966--5983.

\bibitem{NowTecBog19}
Nowak,~A.;\ \ Tecmer,~P.;\ \ Boguslawski,~K. Assessing the accuracy of simplified coupled cluster methods for electronic excited states in f0 actinide compounds. \textit{Phys.\ Chem.\ Chem.\ Phys.} \textbf{2019,} \textsl{21,} 19039--19053.

\bibitem{BulHenScu15}
Bulik,~I.~W.;\ \ Henderson,~T.~M.;\ \ Scuseria,~G.~E. Can single-reference coupled cluster theory describe static correlation? \textit{J.~Chem.\ Theory Comput.} \textbf{2015,} \textsl{11,} 3171--3179.

\bibitem{GomHenScu16}
Gomez,~J.~A.;\ \ Henderson,~T.~M.;\ \ Scuseria,~G.~E. Recoupling the singlet- and triplet-pairing channels in single-reference coupled cluster theory. \textit{J.~Chem.\ Phys.} \textbf{2016,} \textsl{145,} 134103:1--7.

\bibitem{ScuHenBul13}
Scuseria,~G.~E.;\ \ Henderson,~T.~M.;\ \ Bulik,~I.~W. Particle-particle and quasiparticle random phase approximations: {C}onnections to coupled cluster theory. \textit{J.~Chem.\ Phys.} \textbf{2013,} \textsl{139,} 104113:1--13.

\bibitem{FucNiqGon05}
Fuchs,~M.;\ \ Niquet,~Y.-M.;\ \ Gonze,~X.;\ \ Burke,~K. Describing static correlation in bond dissociation by {K}ohn--{S}ham density functional theory. \textit{J.~Chem.\ Phys.} \textbf{2005,} \textsl{122,} 094116:1--13.

\bibitem{VanYanYan13}
Van~Aggelen,~H.;\ \ Yang,~Y.;\ \ Yang,~W. Exchange-correlation energy from pairing matrix fluctuation and the particle-particle random-phase approximation. \textit{Phys.\ Rev.~A} \textbf{2013,} \textsl{88,} 030501:1--5.

\bibitem{TahRen19}
Tahir,~M.~N.;\ \ Ren,~X. Comparing particle-particle and particle-hole channels of the random phase approximation. \textit{Phys.\ Rev.~B} \textbf{2019,} \textsl{99,} 195149:1--12.

\bibitem{For22}
Forster,~A. Assessment of the second-order statically screened exchange correction to the random phase approximation for correlation energies. \textit{J.~Chem.\ Theory Comput.} \textbf{2022,} \textsl{18,} 5948--5965.

\bibitem{YanvanYan13}
Yang,~Y.;\ \ {van Aggelen},~H.;\ \ Yang,~W. Double, {Rydberg} and charge transfer excitations from pairing matrix fluctuation and particle-particle random phase approximation. \textit{J.~Chem.\ Phys.} \textbf{2013,} \textsl{139,} 224105:1--6.

\bibitem{YanvanSte13}
Yang,~Y.;\ \ van Aggelen,~H.;\ \ Steinmann,~S.~N.;\ \ Peng,~D.;\ \ Yang,~W. Benchmark tests and spin adaptation for the particle-particle random phase approximation. \textit{J.~Chem.\ Phys.} \textbf{2013,} \textsl{139,} 174110:1--10.

\bibitem{SheAggYan14}
Shenvi,~N.;\ \ Van~Aggelen,~H.;\ \ Yang,~Y.;\ \ Yang,~W. Tensor hypercontracted pp{RPA}: Reducing the cost of the particle-particle random phase approximation from {O}(r$^6$) to {O}(r$^4$). \textit{J.~Chem.\ Phys.} \textbf{2014,} \textsl{141,} 024119:1--7.

\bibitem{VanYanYan14}
van Aggelen,~H.;\ \ Yang,~Y.;\ \ Yang,~W. Exchange-correlation energy from pairing matrix fluctuation and the particle-particle random phase approximation. \textit{J.~Chem.\ Phys.} \textbf{2014,} \textsl{140,} 18A511:1--11.

\bibitem{YanPenLu14}
Yang,~Y.;\ \ Peng,~D.;\ \ Lu,~J.;\ \ Yang,~W. Excitation energies from particle-particle random phase approximation: {Davidson} algorithm and benchmark studies. \textit{J.~Chem.\ Phys.} \textbf{2014,} \textsl{141,} 124104:1--10.

\bibitem{LiJinYu24a}
Li,~J.;\ \ Jin,~Y.;\ \ Yu,~J.;\ \ Yang,~W.;\ \ Zhu,~T. Particle--particle random phase approximation for predicting correlated excited states of point defects. \textit{J.~Chem.\ Theory Comput.} \textbf{2024,} \textsl{20,} 7979--7989.

\bibitem{LiJinYu24b}
Li,~J.;\ \ Jin,~Y.;\ \ Yu,~J.;\ \ Yang,~W.;\ \ Zhu,~T. Accurate excitation energies of point defects from fast particle--particle random phase approximation calculations. \textit{J.~Phys.\ Chem.\ Lett.} \textbf{2024,} \textsl{15,} 2757--2764.

\bibitem{MarRomLoo24}
Marie,~A.;\ \ Romaniello,~P.;\ \ Loos,~P.-F. Anomalous propagators and the particle-particle channel: Hedin's equations. \textit{Phys.\ Rev.~B} \textbf{2024,} \textsl{110,} 115155.

\bibitem{YuLiZhu25}
Yu,~J.;\ \ Li,~J.;\ \ Zhu,~T.;\ \ Yang,~W. Accurate and efficient prediction of double excitation energies using the particle--particle random phase approximation. \textit{J.~Chem.\ Phys.} \textbf{2025,} \textsl{162,} 094101:1--9.

\bibitem{Tau08}
Taube,~A.~G. \textit{Developments in coupled-cluster theory gradients and potential energy surfaces,} Thesis, University of Florida, 2008.

\bibitem{TauBar09}
Taube,~A.~G.;\ \ Bartlett,~R.~J. Rethinking linearized coupled-cluster theory. \textit{J.~Chem.\ Phys.} \textbf{2009,} \textsl{130,} 144112:1--14.

\bibitem{BarMusLot10}
Bartlett,~R.~J.;\ \ Musia{\l},~M.;\ \ Lotrich,~V.;\ \ Ku{\'s},~T.  The yearn to be hermitian.   In  \textit{Recent Progress in Coupled Cluster Methods: Theory and Applications}; Springer: Dordrecht, NL, 2010.

\bibitem{JanPal80}
Jankowski,~K.;\ \ Paldus,~J. Applicability of coupled-pair theories to quasidegenerate electronic states: A model study. \textit{Int.~J. Quantum Chem.} \textbf{1980,} \textsl{18,} 1243--1269.

\bibitem{Car25}
{Carter}-{Fenk},~K. Diagrammatic simplification of linearized coupled cluster theory. \textit{J.~Phys.\ Chem.~A} \textbf{2025,} \textsl{129,} 7251--7260.

\bibitem{LotBar11}
Lotrich,~V.;\ \ Bartlett,~R.~J. External coupled-cluster perturbation theory: Description and application to weakly interaction dimers. corrections to the random phase approximation. \textit{J.~Chem.\ Phys.} \textbf{2011,} \textsl{134,} 184108:1--8.

\bibitem{ZobSzaSur13}
Zoboki,~T.;\ \ Szabados,~{\'A}.;\ \ Surj{\'a}n,~P.~R. Linearized {Coupled} {Cluster} {Corrections} to {Antisymmetrized} {Product} of {Strongly} {Orthogonal} {Geminals}: {Role} of {Dispersive} {Interactions}. \textit{J.~Chem.\ Theory Comput.} \textbf{2013,} \textsl{9,} 2602--2608.

\bibitem{BogAye15}
Boguslawski,~K.;\ \ Ayers,~P.~W. Linearized {Coupled} {Cluster} {Correction} on the {Antisymmetric} {Product} of 1-{Reference} {Orbital} {Geminals}. \textit{J.~Chem.\ Theory Comput.} \textbf{2015,} \textsl{11,} 5252--5261.

\bibitem{NowLegBog21}
Nowak,~A.;\ \ Legeza,~{\"O}.;\ \ Boguslawski,~K. Orbital entanglement and correlation from pccd-tailored coupled cluster wave functions. \textit{J.~Chem.\ Phys.} \textbf{2021,} \textsl{154,}.

\bibitem{ChaBogTec23}
Chakraborty,~R.;\ \ Boguslawski,~K.;\ \ Tecmer,~P. Static embedding with pair coupled cluster doubles based methods. \textit{Phys.\ Chem.\ Chem.\ Phys.} \textbf{2023,} \textsl{25,} 25377--25388.

\bibitem{NowBog23}
Nowak,~A.;\ \ Boguslawski,~K. A configuration interaction correction on top of pair coupled cluster doubles. \textit{Phys.\ Chem.\ Chem.\ Phys.} \textbf{2023,} \textsl{25,} 7289--7301.

\bibitem{ShaAla15}
Sharma,~S.;\ \ Alavi,~A. Multireference linearized coupled cluster theory for strongly correlated systems using matrix product states. \textit{J.~Chem.\ Phys.} \textbf{2015,} \textsl{143,} 102815:1--9.

\bibitem{YiChe18}
Yi,~J.;\ \ Chen,~F. Application of multireference linearized coupled-cluster theory to atomic and molecular systems. \textit{J. Theor. Comput. Chem.} \textbf{2018,} \textsl{17,} 1850016.

\bibitem{WaiSucKoh25}
Waigum,~A.;\ \ Suchaneck,~S.;\ \ K{\"o}hn,~A. Simplified multireference coupled-cluster methods: Hybrid approaches with averaged coupled pair theories. \textit{J. Comput. Chem.} \textbf{2025,} \textsl{46,} e70020.

\bibitem{vanGor00}
Van~Voorhis,~T.;\ \ Head-Gordon,~M. Benchmark variational coupled cluster doubles results. \textit{J.~Chem.\ Phys.} \textbf{2000,} \textsl{113,} 8873--8879.

\bibitem{MasHumGru24}
Masios,~N.;\ \ Hummel,~F.;\ \ Gr{\"u}neis,~A.;\ \ Irmler,~A. Investigating the {Basis} {Set} {Convergence} of {Diagrammatically} {Decomposed} {Coupled}-{Cluster} {Correlation} {Energy} {Contributions} for the {Uniform} {Electron} {Gas}. \textit{J.~Chem.\ Theory Comput.} \textbf{2024,} \textsl{20,} 5937--5950.

\bibitem{JahAhmBog25}
Jahani,~S.;\ \ Ahmadkhani,~S.;\ \ Boguslawski,~K.;\ \ Tecmer,~P. Simple and efficient computational strategies for calculating orbital energies and pair-orbital energies from p{CCD}-based methods. \textit{J.~Chem.\ Phys.} \textbf{2025,} \textsl{162,} 184110:1--13.

\bibitem{CarHea23}
{Carter-Fenk},~K.;\ \ {Head-Gordon},~M. Repartitioned {Brillouin-Wigner} perturbation theory with a size-consistent second-order correlation energy. \textit{J.~Chem.\ Phys.} \textbf{2023,} \textsl{158,} 234108:1--14.

\bibitem{SheHenScu14}
Shepherd,~J.~J.;\ \ Henderson,~T.~M.;\ \ Scuseria,~G.~E. Range-separated {Brueckner} coupled cluster doubles theory. \textit{Phys.\ Rev.\ Lett.} \textbf{2014,} \textsl{112,} 133002:1--5.

\bibitem{QCHEM5}
Epifanovsky,~E. \textit{et al.}\  Software for the frontiers of quantum chemistry: {An} overview of developments in the {Q}-{Chem}~5 package. \textit{J.~Chem.\ Phys.} \textbf{2021,} \textsl{155,} 084801:1--59.

\bibitem{PySCF1}
Sun,~Q.;\ \ Berkelbach,~T.~C.;\ \ Blunt,~N.~S.;\ \ Booth,~G.~H.;\ \ Guo,~S.;\ \ Li,~Z.;\ \ Liu,~J.;\ \ {McClain},~J.~D.;\ \ Sayfutyarova,~E.~R.;\ \ Sharma,~S.;\ \ Wouters,~S.;\ \ Chan,~G. K.~L. {PySCF}: {The} {Python}-based simulations of chemistry framework. \textit{WIREs Comput.\ Mol.\ Sci.} \textbf{2017,} \textsl{8,} e1340.

\bibitem{PySCF2}
Sun,~Q. \textit{et al.}\  Recent developments in the {\sc pyscf} program package. \textit{J.~Chem.\ Phys.} \textbf{2020,} \textsl{153,} 024109:1--20.

\bibitem{Dun89}
{Dunning, Jr.},~T.~H. Gaussian basis sets for use in correlated molecular calculations. {I}. {The} atoms boron through neon and hydrogen. \textit{J.~Chem.\ Phys.} \textbf{1989,} \textsl{90,} 1007--1023.

\bibitem{KenDunHar92}
Kendall,~R.~A.;\ \ {Dunning, Jr.},~T.~H.;\ \ Harrison,~R.~J. Electron affinities of the first-row atoms revisited. {S}ystematic basis sets and wave functions. \textit{J.~Chem.\ Phys.} \textbf{1992,} \textsl{96,} 6796--6806.

\bibitem{WooDun93}
Woon,~D.~E.;\ \ {Dunning, Jr.},~T.~H. Gaussian basis sets for use in correlated molecular calculations. {III}. {The} atoms aluminum through argon. \textit{J.~Chem.\ Phys.} \textbf{1993,} \textsl{98,} 1358--1371.

\bibitem{W4-11}
Karton,~A.;\ \ Daon,~S.;\ \ Martin,~J. M.~L. {W4-11}: {A} high-confidence benchmark dataset for computational thermochemistry derived from first-principles data. \textit{Chem.\ Phys.\ Lett.} \textbf{2011,} \textsl{510,} 165--178.

\bibitem{KnoAndAmo91}
Knowles,~P.~J.;\ \ Andrews,~J.~S.;\ \ Amos,~R.~D.;\ \ Handy,~N.~C.;\ \ Pople,~J.~A. Restricted {M{\o}ller-Plesset} theory for open-shell molecules. \textit{Chem.\ Phys.\ Lett.} \textbf{1991,} \textsl{186,} 130--136.

\bibitem{HelKloKoc97}
Helgaker,~T.;\ \ Klopper,~W.;\ \ Koch,~H.;\ \ Noga,~J. Basis-set convergence of correlated calculations on water. \textit{J.~Chem.\ Phys.} \textbf{1997,} \textsl{106,} 9639--9646.

\bibitem{def2}
Weigend,~F.;\ \ Ahlrichs,~R. Balanced basis sets of split valence, triple zeta valence and quadruple zeta valence quality for {H} to {Rn}: {D}esign and assessment of accuracy. \textit{Phys.\ Chem.\ Chem.\ Phys.} \textbf{2005,} \textsl{7,} 3297--3305.

\bibitem{def2aug}
Rappoport,~D.;\ \ Furche,~F. Property-optimized {G}aussian basis sets for molecular response calculations. \textit{J.~Chem.\ Phys.} \textbf{2010,} \textsl{133,} 134105:1--11.

\bibitem{MRCC}
Mester,~D.;\ \ Nagy,~P.~R.;\ \ Cs{\'o}ka,~J.;\ \ {Gyevi}-{Nagy},~L.;\ \ {Bern{\'a}t}-{Szab{\'o}},~P.;\ \ Horv{\'a}th,~R.~A.;\ \ Petrov,~K.;\ \ H{\'e}gely,~B.;\ \ Lad{\'o}czki,~B.;\ \ Samu,~G.;\ \ L{\H{o}}rincz,~B.~D.;\ \ K{\'a}llay,~M. Overview of developments in the {MRCC} program system. \textit{J.~Phys.\ Chem.~A} \textbf{2025,} \textsl{129,} 2086--2107.

\bibitem{HaiTubLev19}
Hait,~D.;\ \ Tubman,~N.~M.;\ \ Levine,~D.~S.;\ \ Whaley,~K.~B.;\ \ {Head-Gordon},~M. What levels of coupled cluster theory are appropriate for transition metal systems? {A} study using near-exact quantum chemical values for 3d transition metal binary compounds. \textit{J.~Chem.\ Theory Comput.} \textbf{2019,} \textsl{15,} 5370--5385.

\bibitem{LanKhiDol10}
Landau,~A.;\ \ Khistyaev,~K.;\ \ Dolgikh,~S.;\ \ Krylov,~A.~I. Frozen natural orbitals for ionized states within equation-of-motion coupled-cluster formalism. \textit{J.~Chem.\ Phys.} \textbf{2010,} \textsl{132,} 014109:1--13.

\bibitem{PokIzmKry10}
Pokhilko,~P.;\ \ Izmodenov,~D.;\ \ Krylov,~A.~I. Extension of frozen natural orbital approximation to open-shell references: Theory, implementation, and application to single-molecule magnets. \textit{J.~Chem.\ Phys.} \textbf{2020,} \textsl{152,} 034105:1--13.

\bibitem{GraHer24}
Gray,~M.;\ \ Herbert,~J.~M. Assessing the domain-based local pair natural orbital ({DLPNO}) approximation for non-covalent interactions in sizable supramolecular complexes. \textit{J.~Chem.\ Phys.} \textbf{2024,} \textsl{161,} 054114:1--19.

\bibitem{SteAgaBha25}
Stevenson,~J.~M. \textit{et al.}\  Evidence for significant multi-{L}i$^+$ clustering in common lithium-ion battery electrolytes.  \textbf{2025,} .

\bibitem{IliSau07}
Ilia{\v s},~M.;\ \ Saue,~T. An infinite-order two-component relativistic {Hamiltonian} by a simple one-step transformation. \textit{J.~Chem.\ Phys.} \textbf{2007,} \textsl{126,} 064102.

\bibitem{LiuPen09}
Liu,~W.;\ \ Peng,~D. Exact two-component {Hamiltonians} revisited. \textit{J.~Chem.\ Phys.} \textbf{2009,} \textsl{131,} 031104.

\bibitem{Sau11}
Saue,~T. Relativistic {Hamiltonians} for {Chemistry}: {A} {Primer}. \textit{ChemPhysChem} \textbf{2011,} \textsl{12,} 3077--3094.

\bibitem{LiXiaLiu12}
Li,~Z.;\ \ Xiao,~Y.;\ \ Liu,~W. On the spin separation of algebraic two-component relativistic {Hamiltonians}. \textit{J.~Chem.\ Phys.} \textbf{2012,} \textsl{137,} 154114.

\bibitem{HalHelJor99}
Halkier,~A.;\ \ Helgaker,~T.;\ \ J{\o}rgensen,~P.;\ \ Klopper,~W.;\ \ Olsen,~J. Basis-set convergence of the energy in molecular {H}artree--{Fock} calculations. \textit{Chem.\ Phys.\ Lett.} \textbf{1999,} \textsl{302,} 437--446.

\bibitem{Tru98}
Truong,~T.~N. Quantum modelling of reactions in solution: {An} overview of the dielectric continuum methodology. \textit{Int.\ Rev.\ Phys.\ Chem.} \textbf{1998,} \textsl{17,} 525--546.

\bibitem{NeeVal11}
Neese,~F.;\ \ Valeev,~E.~F. Revisiting the atomic natural orbital approach for basis sets: {Robust} systematic basis sets for explicitly correlated and conventional correlated \textit{ab initio} methods? \textit{J.~Chem.\ Theory Comput.} \textbf{2011,} \textsl{7,} 33--43.

\bibitem{Wei06}
Weigend,~F. Accurate {Coulomb}-fitting basis sets for {H} to {Rn}. \textit{Phys.\ Chem.\ Chem.\ Phys.} \textbf{2006,} \textsl{8,} 1057--1065.

\bibitem{FurPer06}
Furche,~F.;\ \ Perdew,~J.~P. The performance of semilocal and hybrid density functionals in 3d transition-metal chemistry. \textit{J.~Chem.\ Phys.} \textbf{2006,} \textsl{124,} 044103:1--27.

\bibitem{Abdul}
Zamani,~A.~Y.;\ \ Zulueta,~B.;\ \ Ricciuti,~A.~M.;\ \ Keith,~J.~A.;\ \ Carter-Fenk,~K. {K}ohn-{S}ham density encoding rescues coupled cluster theory for strongly correlated molecules. \textit{arXiv preprint https://arxiv.org/abs/2602.06149} \textbf{2026,} .

\bibitem{HolGil11}
Hollet,~J.~W.;\ \ Gill,~P. M.~W. The two faces of static correlation. \textit{J.~Chem.\ Phys.} \textbf{2011,} \textsl{134,} 114111:1--5.

\bibitem{KrySheByr98}
Krylov,~A.~I.;\ \ Sherrill,~C.~D.;\ \ Byrd,~E. F.~C.;\ \ {Head-Gordon},~M. Size-consistent wave functions for nondynamical correlation energy: {The} valence active space optimized orbital coupled-cluster doubles model. \textit{J.~Chem.\ Phys.} \textbf{1998,} \textsl{109,} 10669--10678.

\bibitem{MusBar05}
Musia{\l},~M.;\ \ Bartlett,~R.~J. Critical comparison of various connected quadruple excitation approximations in the coupled-cluster treatment of bond breaking. \textit{J.~Chem.\ Phys.} \textbf{2005,} \textsl{122,} 224102:1--9.

\bibitem{DroHon57}
Drowart,~J.;\ \ Honig,~R. A mass spectrometric method for the determination of dissociation energies of diatomic molecules. \textit{J.~Phys.\ Chem.} \textbf{1957,} \textsl{61,} 980--985.

\bibitem{StaLipMag89}
Stanton,~J.~F.;\ \ Lipscomb,~W.~N.;\ \ Magers,~D.~H.;\ \ Bartlett,~R.~J. Highly correlated single-reference studies of the {O}$_3$ potential surface. i. effects of high order excitations on the equilibrium structure and harmonic force field of ozone. \textit{J.~Chem.\ Phys.} \textbf{1989,} \textsl{90,} 1077--1082.

\bibitem{LeeScu90}
Lee,~T.~J.;\ \ Scuseria,~G.~E. The vibrational frequencies of ozone. \textit{J.~Chem.\ Phys.} \textbf{1990,} \textsl{93,} 489--494.

\bibitem{Hub63}
Hubbard,~J. Electron correlations in narrow energy bands. \textit{Proc.\ A} \textbf{1963,} \textsl{276,} 238--257.

\bibitem{LeeTay89}
Lee,~T.~J.;\ \ Taylor,~P.~R. A diagnostic for determining the quality of single-reference electron correlation methods. \textit{Int. J. Quantum Chem} \textbf{1989,} \textsl{36,} 199--207.

\bibitem{JiaDeYNat11}
Jiang,~W.;\ \ {DeYonker},~N.~J.;\ \ Wilson,~A.~K. Multireference character for 3d transition-metal-containing molecules. \textit{J.~Chem.\ Theory Comput.} \textbf{2011,} \textsl{8,} 460--468.

\bibitem{RanCar26}
Ransford,~E.~R.;\ \ {Carter}-{Fenk},~K. \textit{ChemRxiv} \textbf{2026,}  pre-print; DOI: 10.26434/chemrxiv.10001586/v1.

\bibitem{Pie25}
Pierce,~K. Toward using matrix-free tensor decompositions to systematically improve approximate tensor-networks. \textit{J.~Chem.\ Theory Comput.} \textbf{2025,} \textsl{21,} 6464–-6481.

\bibitem{SonGonYe25}
Song,~R.;\ \ Gong,~X.;\ \ Ye,~H.-Z. Unphysical solutions in coupled-cluster-based random phase approximation and how to avoid them. \textit{J.~Chem.\ Phys.} \textbf{2025,} \textsl{163,} 161103:1--8.

\end{thebibliography}

\providecommand{\refin}[1]{\\ \textbf{Referenced in:} #1}

\clearpage
For Table of Contents Only
\\
\begin{center}
TOC Graphic\\
\fig{1.0}{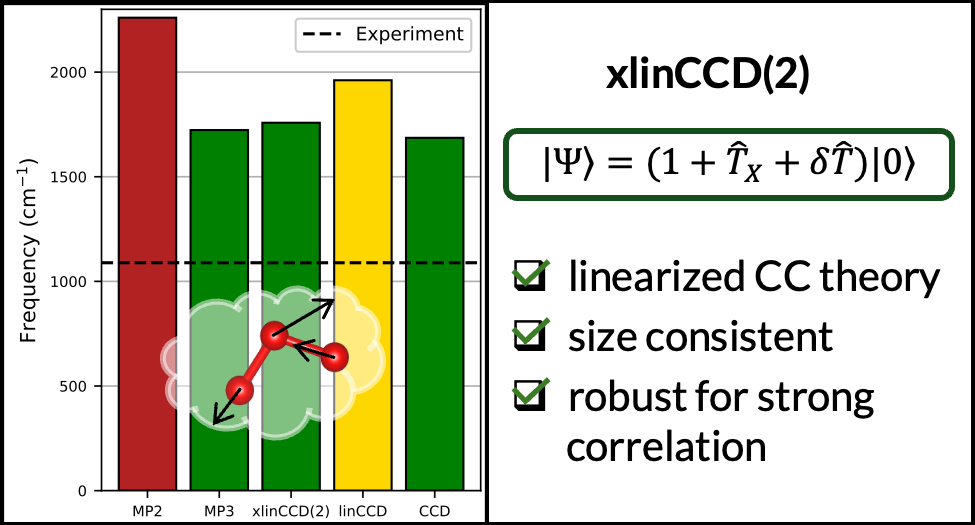}
\end{center}
\end{document}


\title{
Supporting Information for 
\\
The Dose Makes the Poison: Perturbative Steps Toward the Ultimate Linearized Coupled Cluster Method
}

\author{Sylvia J. Bintrim}
\author{Ella R. Ransford}
\author{
    Kevin Carter-Fenk*\footnote[0]{*kay.carter-fenk@pitt.edu}
}
\affiliation{
	Department of Chemistry, University of Pittsburgh, Pittsburgh, Pennsylvania 15218, USA
}

\date{\today}

\maketitle
\thispagestyle{plain}
\section{Mosaic-type terms in xlinCCD(2)}
\begin{figure}[H]
    \centering
    \fig{1.0}{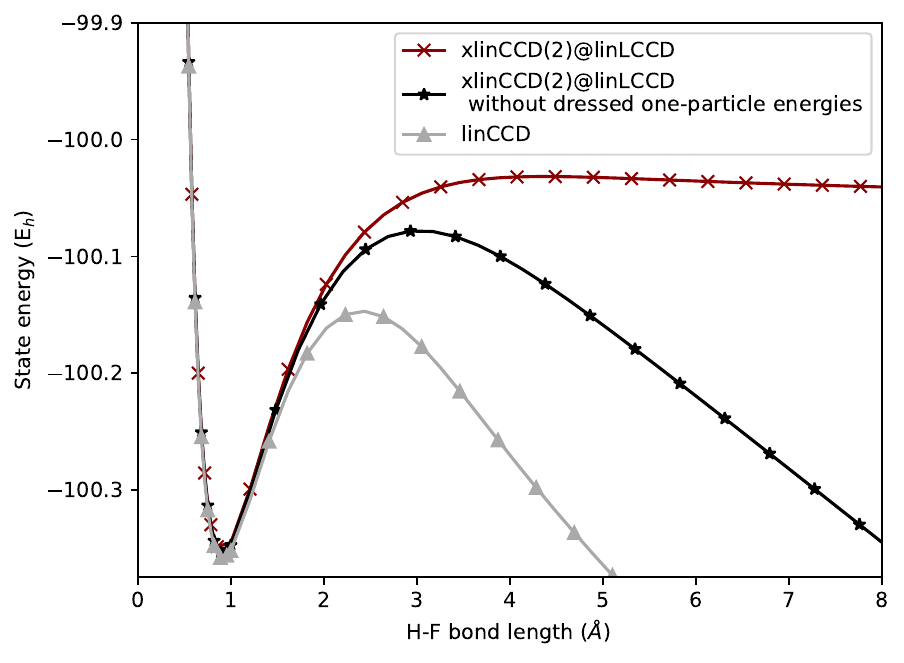}
    \caption{Ground state dissociation curves of FH in the aug-cc-pVTZ basis.}
    \label{fig:H0_choice}
\end{figure}
Figure~\ref{fig:H0_choice} shows the importance of employing dressed one-particle energies 
in xlinCCD(2) for bond dissociation. Without this choice of $\hat{H}_0$ resulting in mosaic-type terms in the xlinCCD(2) equation, xlinCCD(2) still performs better than linCCD but also eventually diverges downward for bond dissociation.

\clearpage\pagebreak
\section{Minimal Basis H$_2$}
\begin{figure}[h!!]
    \centering
    \fig{1.0}{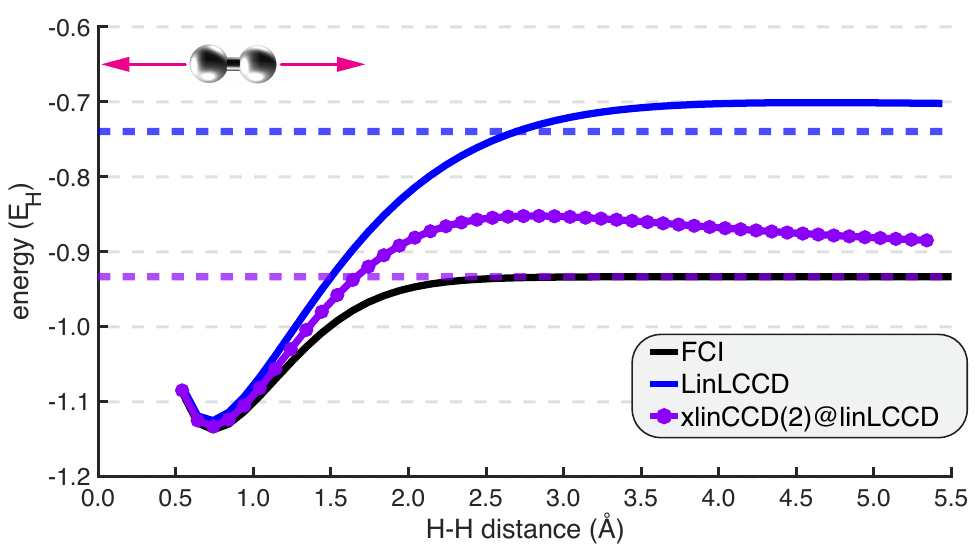}
    \caption{Restricted linLCCD and xlinCCD(2)@linLCCD potential energy surfaces for H$_2$ bond dissociation in the STO-3G basis compared with the (exact) CCSD solution. Dashed lines are estimates of the asymptotic limits calculated at 10,000~\AA.}
    \label{fig:H2minbasis}
\end{figure}

Figure~\ref{fig:H2minbasis} shows the bond dissociation potential energy surfaces obtained for minimal basis H$_2$. Critically, the xlinCCD(2)@linLCCD method produces the exact dissociation limit for this 2-electrons in 2-orbitals case. Other improvements in relative energy are seen across the entire potential surface when using xlinCCD(2)@linLCCD.

\section{W4-11 Thermochemistry}
\begin{figure}
    \centering
    \fig{1.0}{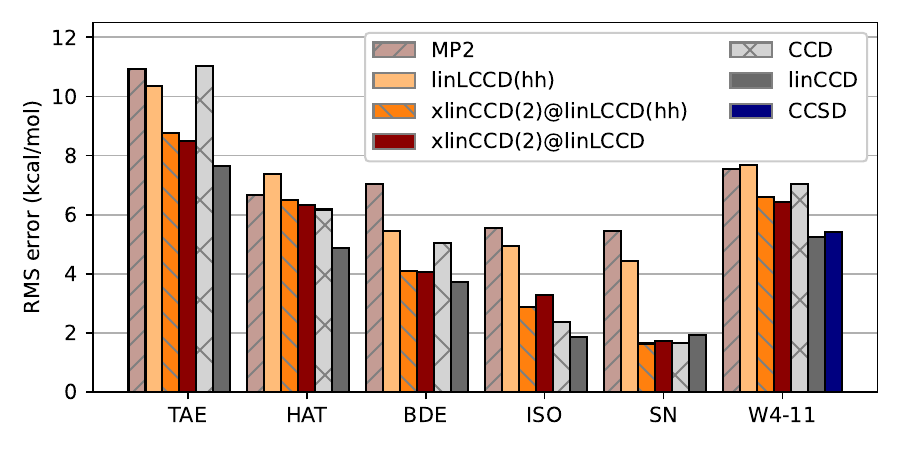}
    \caption{Root-mean-square errors (kcal/mol) produced by various wave function methods applied to subsets of thermochemical processes within W4-11: non-multireference total atomization energies (TAE), non-multireference heavy-atom transfer (HAT), non-multireference bond-dissociation energies (BDE), isomerization (ISO), and nucleophilic substitution (SN) energies. 
    MP2 data is taken from Ref.~\citen{CarSheHea23} and CCSD data from Ref.~\citen{LeePhaRei22}.}
    \label{fig:w411}
\end{figure}
For the 83 non-multi-reference bond dissociation processes 
in Figure~\ref{fig:w411}, we see that none of the wave function methods assessed here provide BDEs within chemical accuracy (2 kcal/mol). Both varieties of xlinCCD(2) and linCCD gave somewhat more accurate BDEs than CCD. Together with our qualitatively correct dissociation curves, these results suggest that xlinCCD(2) provides results on par with CCD for bond-breaking processes.

MP2, linLCCD(hh), xlinCCD(2), and CCD provide overall W4-11 thermochemical data with substantial root-mean-square errors in the range of 6-8 kcal/mol. Interestingly, linCCD generally performs slightly better than the other methods. We conclude that xlinCCD(2) typically provides results of a quality comparable to those of linCCD or CCD for these weakly-correlated systems. Like other simplified or regularized methods tailored to the strong correlation problem,\cite{WanSheHea26} the actual benefits of the xlinCCD(2) approach become more apparent for more challenging problems like bond dissociation curves.
\section{BDEs of first-row transition metal heteronuclear diatomics}
\begin{table}[H]
\caption{Mean absolute
errors (kcal/mol) in BDEs for first-row transition metal heteronuclear diatomics, computed using CC methods. Reference bond lengths, spin states, and BDEs are from Ref.s~\citen{FurPer06} and \citen{Abdul}. Spin-orbit corrections from Ref.~\citen{Abdul} were applied.}
\label{tab:tm_diatomics}
\begin{tabular}{l|l|l|l|l|l|l|l}
\hline\hline
 
&linLCCD(hh)
&linLCCD
&\makecell{xlinCCD(2)@\\linLCCD(hh)}&\makecell{xlinCCD(2)@\\linLCCD} 
&linCCD
&CCD & CCSD\\
\hline
\hline
hydrides & \cellcolor{green!30}3.71 & \cellcolor{yellow!30}5.14	&	\cellcolor{orange!30}5.70	&	\cellcolor{yellow!30}5.23	&	\cellcolor{orange!30}5.71	&	\cellcolor{GreenYellow!30}5.01	&	\cellcolor{red!30}6.44	\\
\hline
chlorides & \cellcolor{red!30}7.81 &  \cellcolor{orange!30}6.10	&	\cellcolor{yellow!30}5.20	&	\cellcolor{yellow!30}5.20	&	\cellcolor{green!30}4.17	&	\cellcolor{GreenYellow!30}4.47	&	\cellcolor{yellow!30}5.31\\
\hline
oxides & \cellcolor{green!30}11.01 & \cellcolor{red!30}25.05 &\cellcolor{yellow!30}19.61&\cellcolor{GreenYellow!30}16.96 & \cellcolor{yellow!30}19.28&\cellcolor{orange!30}22.03 & \cellcolor{GreenYellow!30}14.54
\\
\hline
 all three & \cellcolor{green!30}7.49 & \cellcolor{red!30}12.41 & 
 \cellcolor{yellow!30}10.20 & \cellcolor{GreenYellow!30}9.12 & \cellcolor{yellow!30}9.98 & \cellcolor{orange!30}10.65 & \cellcolor{GreenYellow!30}8.78 
\\
\hline\hline
\end{tabular}
\end{table}

\section{One-shot reformulation of xlinCCD(2)@linLCCD(hh)}
Although we employ only the na{\"i}ve implementation in this paper, change-of-basis implementations of linLCCD(hh) and xlinCCD(2)@linLCCD(hh) that are isomorphic with MP2 may reduce the methods' execution times. As detailed in Ref.~\citen{Car25}, we can perform a memory-efficient linLCCD(hh) calculation via (1) an initial 
diagonalization of dressed Fock elements in the hole-hole space and (2) subsequent, one-shot, $N^4$-scaling perturbative step if we employ density fitting/resolution of the identity for the two-electron integrals.\cite{Car25} 

A similar re-formulation of xlinCCD(2) is possible on top of the linLCCD(hh) reference.
First, we can simultaneously diagonalize the dressed one-particle energies $X_i^j$ and $X_b^a$ $(X_i^j \oplus X_b^a)$
:
\begin{subequations}
\begin{equation}
X_b^a=f_b^a-\frac{1}{2}t_{\text{X}mn}^{\phantom{\text{X}}ae}v^{mn}_{be}
\rightarrow \tilde{X}_b
\end{equation}
\begin{equation}
X_i^j=f_i^j+\frac{1}{2}t_{\text{X}im}^{\phantom{\text{X}}ef}v^{jm}_{ef}
\rightarrow \tilde{X}_j
\end{equation}
\end{subequations}
Note that for the xlinCCD(2) correction on top of linLCCD(hh), we have
\begin{equation}
\begin{split}
X_{ij}^{ab}&= \frac{1}{2}t_{\text{X}ij}^{\phantom{\text{X}}ef}v^{ab}_{ef}+\mathcal{P}_{ij}\mathcal{P}_{ab}(t_{\text{X}im}^{\phantom{\text{X}}ae}v^{mb}_{ej})
\end{split}
\end{equation}
as we must include the linear, particle-particle ladder term as well as the linear ring/crossed ring terms in the perturbative step.

Next, we can perform the one-shot, $N^6$-scaling xlinCCD(2) correction 
in the basis of eigenvectors of $X_i^j \oplus X_b^a$, denoted by tilde symbols:
\begin{equation}
\delta t_{ij}^{ab}=\frac{\frac{1}{2}\tilde{t}_{\text{X}ij}^{\phantom{\text{X}}ef}\tilde{v}^{ab}_{ef}+\mathcal{P}_{ij}\mathcal{P}_{ab}(\tilde{t}_{Xim}^{\:\:\:ae}\tilde{v}^{mb}_{ej})}{
2\tilde{X}_b
-2\tilde{X}_j}
\end{equation}
While CCD, linCCD, and xlinCCD(2)@linLCCD require iterative solution, this one-shot reformulation of xlinCCD(2)\allowbreak@linLCCD(hh) may have lower computational execution time.

\section{TinySpins25 Data Set}
The TinySpins25 data set contains estimates of the lowest-energy singlet-triplet gap for 25 heteronuclear transition-metal diatomics.
We optimized the bond lengths at the $\omega$B97M-V/Def2-TZVPP level of theory.
The CCSDT(Q)$_\Lambda$ calculations were extrapolated
to the complete basis set limit with a two-point
extrapolation scheme using the Def2-SVPD and Def2-TZVPPD basis
sets. We applied the same frozen core scheme as Ref.~\citen{HaiTubLev19} for 3d metals and applied
the pseudopotential that was optimized for the Karlsruhe basis
sets for all 4d and 5d metal atoms. When pseudopotentials were applied, only core orbitals on the non-metal atom were
frozen. We tested this CBS extrapolation scheme by comparing
our results for RuC with the high-level (aug-cc-pV5Z-PP) multireference configuration interaction (MRCI) data from
Ref.~\citen{TzeKar20}. Their MRCI/aug-cc-pV5Z-PP data are in excellent agreement with experiment and their estimated gap is 2.1~kcal/mol. Our CCSDT(Q)$_\Lambda$/CBS result is
2.0~kcal/mol, which is in near-perfect agreement with large-basis MRCI. Our triple-$\zeta$ estimates for the gaps of ScH, ScF, CuH, CuF, CuCl, ZnO, and ZnS (the subset of TinySpins25 that overlaps
with Quest~\#8)\cite{JacKosGam23} are also within 1~kcal/mol
of the aug-cc-pVTZ Quest~\#8 theoretical best estimates (which often equates to FCI). Overall, we estimate that the accuracy of our predicted gaps is less than 1~kcal/mol from the true non-relativistic values.
It is our intention to supply the community with a non-relativistic estimate of all of these gaps (save the relativistic effects that are inherent to the Karlsruhe pseudopotentials), so TinySpins25 does not make use of
scalar nor vector relativistic corrections.

Finally, we note that no
$T_1$ diagnostic on any system in TinySpins25 exceeds 0.05, implying that this may be a good test set to benchmark density functionals due to its single-reference character. However, further assessments of the multiconfigurational nature of these small-gap systems would be necessary to make that determination.


\providecommand{\refin}[1]{\\ \textbf{Referenced in:} #1}